\documentclass[printer]{aa}
\usepackage{graphicx}
\usepackage{multirow}
\usepackage[colorlinks=true,linkcolor=blue]{hyperref}
\usepackage{threeparttable}
\usepackage{txfonts}
\begin{document}

            \title{Evidence of the $Gaia$--VLBI position differences being related to radio source structure}

  \author{Ming H. Xu
          \inst{1,2,4}
          \and
          Susanne Lunz\inst{3}
          \and
          James M. Anderson\inst{4}
          \and
          Tuomas Savolainen\inst{1,2}
          \and
          Nataliya Zubko\inst{6}
          \and 
          Harald Schuh\inst{4,3}
          }

   \institute{Aalto University Mets\"{a}hovi Radio Observatory, Mets\"{a}hovintie 114, 02540 Kylm\"{a}l\"{a}, Finland; \email{minghui.xu@aalto.fi}
         \and
             Aalto University Department of Electronics and Nanoengineering, PL15500, FI-00076 Aalto, Finland
             \and
DeutschesGeoForschungsZentrum (GFZ), Potsdam, Telegrafenberg, 14473 Potsdam, Germany
	\and
Institute of Geodesy and Geoinformation Science, Technische Universit\"{a}t Berlin, Stra{\ss}e des 17. Juni 135, 10623, Berlin, Germany
\and
Finnish Geospatial Research Institute, Geodeetinrinne 2, FIN-02430 Masala, Finland}

   \date{Received ***; accepted ***}

\titlerunning{Position differences between $Gaia$ and VLBI}

 
  \abstract
   {We report the relationship between 
   the $Gaia$--VLBI position differences and the magnitudes of source structure effects in VLBI observations.}
   {Because the $Gaia$--VLBI position differences are statistically significant for a considerable number of common sources, we attempt to discuss and explain these position differences based on VLBI observations and available source images at cm-wavelengths. }
   {Based on the derived closure amplitude root-mean-square (CARMS), which quantifies the magnitudes 
   of source structure effects in the VLBI observations used for building the third realization of the International Celestial Reference Frame, the arc lengths and normalized arc lengths of the position differences are examined in detail. 
  The radio jet directions and the directions of the $Gaia$--VLBI position differences are investigated for a small sample
  of sources.
   }
   {
   Both the arc lengths and normalized arc lengths of the $Gaia$ and VLBI positions are found to increase with the CARMS values. The majority of the sources with statistically significant position differences are associated with the sources having extended structure. Radio source structure is the one of the major factors of these position differences, and it can be the dominate factor for a number of sources. The vectors of the $Gaia$ and VLBI position differences are parallel to the radio-jet directions, which is confirmed with stronger evidence.   }
   {}


   \maketitle
%

\section{Introduction}
The International Celestial Reference Frame (ICRF) was adopted 
as the Fundamental Celestial Reference Frame for astronomy in January 1998
by the International Astronomical Union (IAU) \citep{1998AJ....116..516M}. 
The ICRF is realized by the positions of distant radio sources, mostly active
galactic nuclei (AGNs), based on the astrometric/geodetic very-long-baseline interferometry (VLBI) observations
coordinated by the International VLBI Service for Geodesy and Astrometry
\citep[IVS;][please also refer to the IVS
website\footnote{\url{https://ivscc.gsfc.nasa.gov/index.html}}]{schuh2012jg,nothnagel2017jg}, and relies on
the VLBI technique for its maintenance and improvement. 
As officially adopted by the IAU in January 2019, the third realization of the ICRF \citep[ICRF3;][]{ICRF3_2020} 
is established based on 40 years of VLBI observations and, for the first time, at three different radio frequencies independently. The radio source positions in the ICRF3 have 
accuracies at the sub-milliarcsecond levels.
The European Space Agency 
mission $Gaia$\footnote{\url{https://sci.esa.int/web/gaia}} \citep{2016A&A...595A...2G} has 
released position estimates 
and other astrometric parameters for the celestial
objects with optical $G$ magnitudes < 21 mag based on the observations during the 22 months since July 2014 \citep[DR2;][]{2018A&A...616A...1G}. The color-dependent calibration is possible based on the $Gaia$ DR2 and leads to 
improvements in the astrometric solution thereafter \citep{2020arXiv201203380L}. $Gaia$ Early Data Release 3 \citep[EDR3;][]{2020arXiv201201533G} has made the data available based on the first 34 months of its observations.

A good overall agreement between radio and optical positions was achieved 
for the cross-matched common objects \citep{2016A&A...595A...5M,2018A&A...616A..14G}; 
the median arc length between the source positions from $Gaia$ and VLBI is $\sim$0.5\,milliarcsecond 
(mas) based on the $Gaia$ DR2 and the ICRF3. 
However, the distribution of the arc lengths between radio and optical positions 
normalized by their uncertainties, called normalized arc length hereafter, deviates
from the expected Rayleigh distribution with $\sigma$ = 1. 
The most obvious deviations in that distribution 
are the long tail spreading to very large normalized arc lengths and the significant 
deficit of values in the bins around the expected peak. 
In the $Gaia$ DR1, there were only a few percent of sources with normalized
arc lengths > 3 \citep{2016A&A...595A...5M,2017MNRAS.467L..71P}. In the $Gaia$ DR2, the number of such sources even increases 
to larger than 10 percent \citep{2017MNRAS.471.3775P,2018A&A...616A..14G,2019MNRAS.482.3023P,2019ApJ...873..132M}. By deselecting objects mostly based on the optical properties, \cite{2019ApJ...873..132M} still found 20 percent of sources having normalized arc lengths > 3. The factors causing these position differences
between optical and radio are still unclear, even though there are a variety of
possible astrophysical explanations \citep{2019ApJ...873..132M,2019ApJ...871..143P,2020MNRAS.493L..54K}. For instance,
\cite{2017A&A...598L...1K} and \cite{2019MNRAS.482.3023P} suggested that the main cause of the
position differences is optical structure, the optical jets at the mas scales. Understanding these position 
differences is very important because (1) it will lead to a better 
selection of the common sources for aligning the optical frame 
to the radio frame; (2) the number of 
sources with statistically significant position differences can continue to increase in 
future $Gaia$ data releases, which would allow more and more small position differences be
detected at the 3$\sigma$ confidence level; and (3) the position differences may tell something 
important about the astrophysics of the AGNs.

We examine the position differences between $Gaia$ and VLBI from the radio side. As demonstrated in 
the imaging survey of radio sources \citep{1990A&A...229...51C,1997ApJS..111...95F}, 
the celestial reference frame (CRF) sources commonly have angular 
structure at the mas scales at cm-wavelengths\footnote{Refer to the images of CRF sources at \url{http://www.physics.purdue.edu/astro/MOJAVE/.}}. Source structure is time and frequency dependent, and it is not modeled in the data analysis of building the ICRF3. The position estimates and their uncertainties in the ICRF3 are based on global least-square fitting (LSQ) and are thus not able to characterize the impacts of the systematical position variations over the 40 years due to source structure. For example, the position uncertainties from LSQ are likely 
underestimated in the presence of systematic errors. 
Our previous study has used the same VLBI observations as for the
creation of the ICRF3 to quantify the magnitude of effects of source structure on VLBI observables 
for each individual source \citep{xu2019apjs}. In this paper, we 
apply the results to investigate the relationship between the $Gaia$--VLBI position differences and 
source structure at the cm-wavelengths. We then attempt to explain and discuss 
these position differences based on the radio images from the Monitoring Of Jets in Active galactic 
nuclei with VLBA Experiments \citep[MOJAVE;][]{2018ApJS..234...12L}.

The paper is structured as follows. We introduce in Sect. 2 how the arc lengths
of position differences, the normalized arc lengths, and the quantitative values of 
measuring structure effects are derived. We describe in Sect. 3 the results from 
the examination of arc lengths, normalized arc lengths, optical $G$ magnitudes, and
redshifts with respect to source structure. In Sect. 4, the following topics are discussed: (1) source structure and its quantification; (2) the impact of frequency dependence of source structure; (3) the large position differences that are statistically significant; (4) the magnitudes 
of the position differences; and (5) 
the directions of the position differences. We make the conclusion in Sect. 5.

\section{Data}
\subsection{Source positions from $Gaia$ and VLBI}
We used the right ascension and declination estimates, their uncertainties and the correlations between these two coordinates  
in the ICRF3\footnote{\url{http://hpiers.obspm.fr/icrs-pc/newwww/icrf/icrf3sx.txt}}, 
which contains 4536 sources observed by astrometric/geodetic VLBI at S/X band. 
The median uncertainties of right ascension and declination reported in the ICRF3 
are 0.155\,mas and 0.217\,mas, respectively. We used the $Gaia$ DR2 and EDR3\footnote{\url{https://gea.esac.esa.int/archive/}} 
to get the five astrometric parameters (source position, proper motion, and parallax), their uncertainties, the correlations between them, and the optical magnitude. 

Even though the cross match between radio and optical catalogs basically relies on the
position coincidence, other criteria are needed to reduce the risk of false match.
\cite{2018A&A...616A...2L} applied constraints on the other three astrometric parameters 
and the number of observations, and masked out the region near the Galactic plane, as shown
in Eq. (13) of the publication. \cite{2017MNRAS.467L..71P} used the concept of 
probability of false association as a function of $Gaia$ source density on a 
regular grid and the possible area defined by the positions and the uncertainties 
at radio and optical wavelengths for each potential match. We combined these two methods to 
identify the common objects between the ICRF3 and the $Gaia$ DR2, which gives 
2970 sources \citep[][please refer to the 
poster\footnote{\url{http://www.oan.es/evga2019/EVGA2019_PDF/P310_EVGA2019_Lunz.pdf}}]{2019evga.con_lunz}. 
Based on the $Gaia$ EDR3 and the ICRF3, we identified 3142 common sources, the same number of matched sources as found by the $Gaia$ team in the on-going analysis (Fran\c{c}ois Mignard, private communication).

For each common source, we calculated the arc length between 
the $Gaia$ and VLBI positions, $\rho$, by 
\begin{equation}
\label{rho}
\rho = \sqrt{(\Delta_{\alpha}\cos\delta)^{2}+\Delta_{\delta}^{2}},
\end{equation}
where $\Delta_{\alpha}$ and $\Delta_{\delta}$ are the differences of right ascension and 
declination in the $Gaia$ data and the ICRF3, respectively, and
$\delta$ is the declination. 
The normalized arc length, $X_{\rho}$, is defined and calculated by 
\begin{equation}
\label{X_rho}
X_{\rho} = \rho/\sigma_{\rho},
\end{equation}
where $\sigma_{\rho}$ is the uncertainty of $\rho$ based on the full 2$\times$2 covariance 
matrix, as described in Eqs. (4) and (5) of \cite{2016A&A...595A...5M}. 

To characterize the position uncertainty with a single
value, the semi-major axis of the error ellipse, $\sigma_{\texttt{pos,max}}$, was computed
for both $Gaia$ and VLBI by
\begin{equation}
\label{pos_mas}
\begin{split}
\sigma_{\texttt{pos,max}}^{2} = & \frac{1}{2}\Bigg[ 
 \ (\sigma_{\alpha}\cos\delta)^{2} + \sigma_{\delta}^{2}\\
& + \sqrt{\left((\sigma_{\alpha}\cos\delta)^{2}-\sigma_{\delta}^{2}\right)^{2}+(2\text{C}_{\alpha\delta}\sigma_{\alpha}\cos\delta\sigma_{\delta})^{2}}\ 
\Bigg],
\end{split}
\end{equation}
where $\sigma_{\alpha}$ and $\sigma_{\delta}$ are the uncertainties of right ascension and 
declination, respectively, and $\text{C}_{\alpha\delta}$ is the correlation coefficient 
of the two coordinates.

Based on the $Gaia$ DR2 and the ICRF3, there are 732 sources with $X_{\rho}$ > 3.0; 
for the $Gaia$ EDR3, 804 sources have $X_{\rho}$ > 3.0.

\subsection{Closure amplitude root-mean-square (CARMS)}
We adopted the \emph{log} closure amplitude root-mean-square (CARMS) values from Table 2 in \citet{xu2019apjs} to
quantify the magnitude of source structure effects for each individual source\footnote{The complete table is available 
through the CANFAR data DOI at: \url{https://www.canfar.net/citation/landing?doi=20.0010}.}.

Due to the missing data for calibration and the insensitivity of the parameters 
of geodetic concern, visibility amplitudes from interferometry were not used for most of the 
geodetic VLBI observations. However, they carry valuable information about source angular structure, 
which causes structure effects in group delays up to hundreds of 
picoseconds \citep{1990AJ.....99.1309C,2016AJ....152..151X}.
By forming quadrangles with four baselines, one can get a ratio of the four amplitude
observables to cancel out exactly the station-based errors, which is called closure amplitude 
and provides information about the intrinsic source structure. For an ideal point-like
source, all the baselines will observe the same amplitude within the 
thermal noise, giving closure amplitudes close to unity; for an extended source, 
the closure amplitudes deviate from unity. The CARMS value of a source is defined to be 
the root-mean-square (rms) of \emph{log} closure amplitudes at the X-band based on the 
basic weighting scheme (See Eqs. (2)--(4) and (6)--(8) in \citet{xu2019apjs}). 
In addition to the study in \citet{xu2019apjs}, please also refer to 
its supporting material\footnote{\url{https://www.canfar.net/citation/landing?doi=19.0007}},
where the closure phase and closure amplitude plots are 
available for tens of sources to demonstrate the source structure effects and compare with 
their CARMS values.

The CARMS values are available for 3417 radio sources in the ICRF3 and were derived from 
the astrometric/geodetic VLBI observations from 1979 to 2018, the same dataset 
establishing the ICRF3. They are in the range 0.03--1.63, and the mean and median 
values are 0.31 and 0.24, respectively.
The CARMS values generally classify the CRF sources into three categories:
   \begin{enumerate}
\item CARMS < 0.2 indicates minimum structure;
\item CARMS > 0.3 indicates significant structure;
\item CARMS > 0.4 indicates very extended structure.
\end{enumerate}
The CARMS values were validated by the different source categories in the ICRF catalogs.
For instance, the 39 special handling sources in the second realization of the ICRF \citep{2015AJ....150...58F}, 
which have variations 
in the time series of VLBI position estimates at 
the mas or sub-mas levels, have the median CARMS of 0.60,  
while the median value for the ICRF3 defining sources, used for defining the axis directions 
of the ICRF3, is 0.25. Recently, these CARMS values were used to select radio sources with minimum structure
to assess the quality of group delays in the broadband VLBI system \citep{xu2020jg}.
%
%

For the 3142 common sources from the $Gaia$ EDR3, the CARMS values are available for
2460 sources, 78 percent; the mean and median CARMS 
values are 0.30 and 0.24, respectively, which are at the same
level as those of the 3417 sources. 
We examined the source position estimates 
based on both the $Gaia$ DR2 and EDR3, but we will 
focus on the results from the EDR3 in our study. 

\subsection{Redshift}

We used the Optical Characteristics of Astrometric Radio Sources catalog \citep[OCARS;][]{Malkin_2018} 
to search for the redshifts. The OCARS conveniently provides the redshifts 
for radio sources by collecting them in the literature. Among the 2460 sources, we 
got the redshifts for 2198 sources, $\sim$89 percent. They are in the range 0.01--5.06 with 
mean and median values of 1.28 and 1.18, respectively.

\section{Results}
\subsection{Arc length $\rho$}

We examined the arc lengths $\rho$ between the VLBI and $Gaia$ source position
estimates with respect to the CARMS values. Table \ref{tab1_CARMS} shows the mean and median 
values of $\rho$ and $\sigma_{\rho}$ for different
ranges of CARMS values. The median $\rho$ steadily increases from $\sim$0.4\,mas 
to $\sim$1.3\,mas when the CARMS values increase from 0.4.
The mean $\rho$ increases more significantly from $\sim$0.7\,mas to $\sim$3.7\,mas. 
The median $\rho$ begins to arise when CARMS $\simeq$ 0.6; 
the mean $\rho$ arises significantly, above 1.0\,mas, when CARMS $\simeq$ 0.3. 
We note that in general a smaller CARMS value of a source 
indicates that it causes less structure effects. 

When CARMS < 0.3, the arc lengths $\rho$ have mean values of
$\sim$0.7 mas and median values of $\sim$0.4\,mas, and their uncertainties have mean values
of $\sim$0.4 and median values of $\sim$0.3. It is reasonable to expect that 
these arc lengths will decrease with better uncertainties from $Gaia$ in the near 
future, as happened from the DR2 to the EDR3.
However, when CARMS > 0.6, the arc lengths are larger and statistically significant, and 
they have even better uncertainties than the sources with CARMS < 0.3. 
It is obvious that the sources with extended structure have larger position differences between VLBI and $Gaia$, which are statistically very significant, whereas the sources with minimum structure 
tend to have smaller position differences, which are statistically insignificant.  

The mean $\rho$ is always larger than the median due to a small fraction of sources having considerably larger $\rho$ than the rest of sources in each group. The differences between the mean and median $\rho$ increase with the CARMS values.

\begin{table*}[htbp!]
\caption{Arc lengths $\rho$ and normalized arc lengths $X_{\rho}$ with respect to CARMS}
\label{tab1_CARMS}       
\centering  
\begin{tabular}{crrrcrrrrr}
\hline\noalign{\smallskip}
\multirow{2}{*}{ CARMS} & \multirow{2}{*}{$N_{\texttt{src}}$} & \multicolumn{2}{c}{ $\rho$ [mas]} && \multicolumn{2}{c}{$X_{\rho}$}&& \multicolumn{2}{c}{$\sigma_{\rho}$ [mas]} \\\noalign{\smallskip}\cline{3-4}\cline{6-7}\cline{9-10}\noalign{\smallskip}
&&Mean&Median&&Mean&Median&&Mean&Median\\
\noalign{\smallskip}\hline\noalign{\smallskip}
 < 0.10            & 207 &  0.717 &  0.459&&  1.825 &  1.492 & &  0.382 &  0.304\\
$[$ 0.10 -- 0.20 ) & 724 &  0.710 &  0.448&&  2.135 &  1.600 & &  0.350 &  0.269\\
$[$ 0.20 -- 0.30 ) & 617 &  0.772 &  0.418&&  2.845 &  1.751 & &  0.297 &  0.229\\
$[$ 0.30 -- 0.40 ) & 334 &  1.080 &  0.411&&  3.665 &  1.876 & &  0.287 &  0.218\\
$[$ 0.40 -- 0.50 ) & 220 &  1.347 &  0.452&&  5.425 &  2.543 & &  0.258 &  0.189\\
$[$ 0.50 -- 0.60 ) & 128 &  1.238 &  0.483&&  5.286 &  2.396 & &  0.268 &  0.215\\
$[$ 0.60 -- 0.70 ) &  87 &  1.634 &  0.845&&  6.273 &  3.747 & &  0.324 &  0.252\\
$[$ 0.70 -- 0.80 ) &  59 &  1.536 &  0.769&&  7.918 &  3.842 & &  0.268 &  0.173\\
$[$ 0.80 -- 0.90 ) &  35 &  3.516 &  1.081&& 12.247 &  5.686 & &  0.327 &  0.245\\
$\geq$ 0.90        &  49 & 3.660  &  1.334&& 14.502 &  6.894 && 0.276  &  0.204 \\
\noalign{\smallskip}\hline\noalign{\smallskip}
all               &2460 & 1.012  & 0.459 & & 3.628 &  1.843 & &  0.314 &  0.240\\
\noalign{\smallskip}\hline
\end{tabular}
\end{table*}

\subsection{Normalized arc length $X_{\rho}$}
\begin{figure*}[tbhp!]
   \centering
   \includegraphics[width=0.6\textwidth]{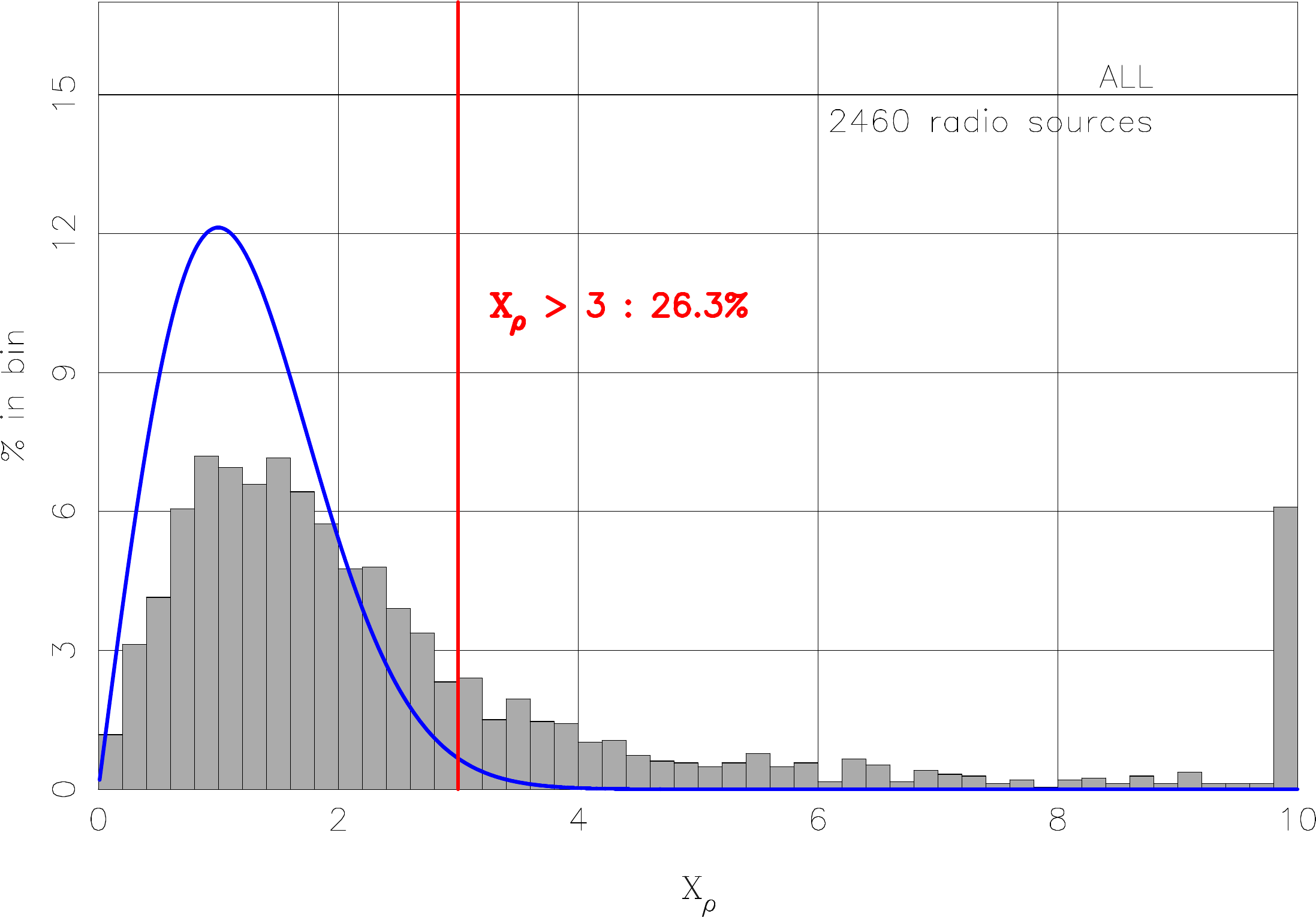}
      \includegraphics[width=0.6\textwidth]{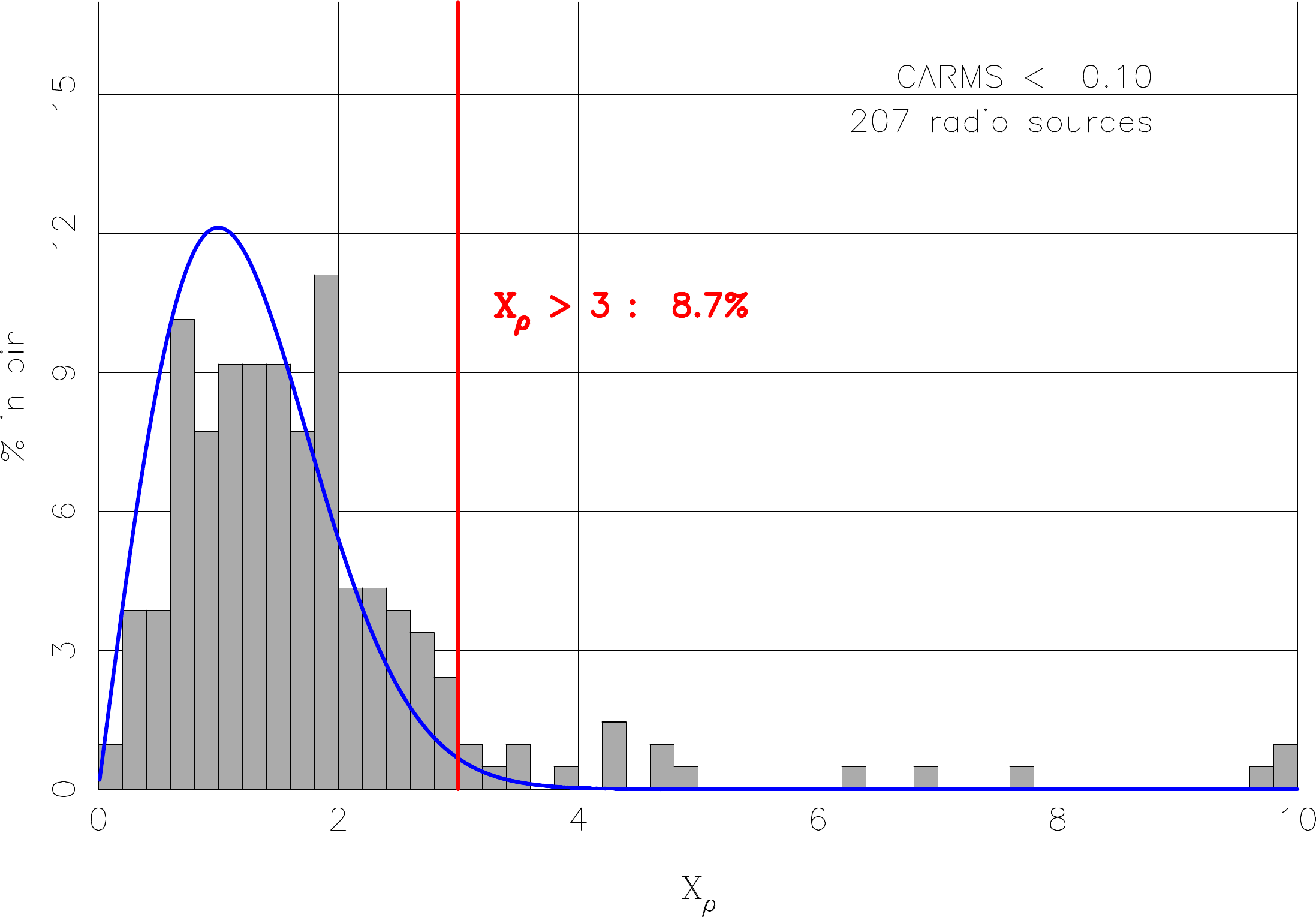}
      \includegraphics[width=0.6\textwidth]{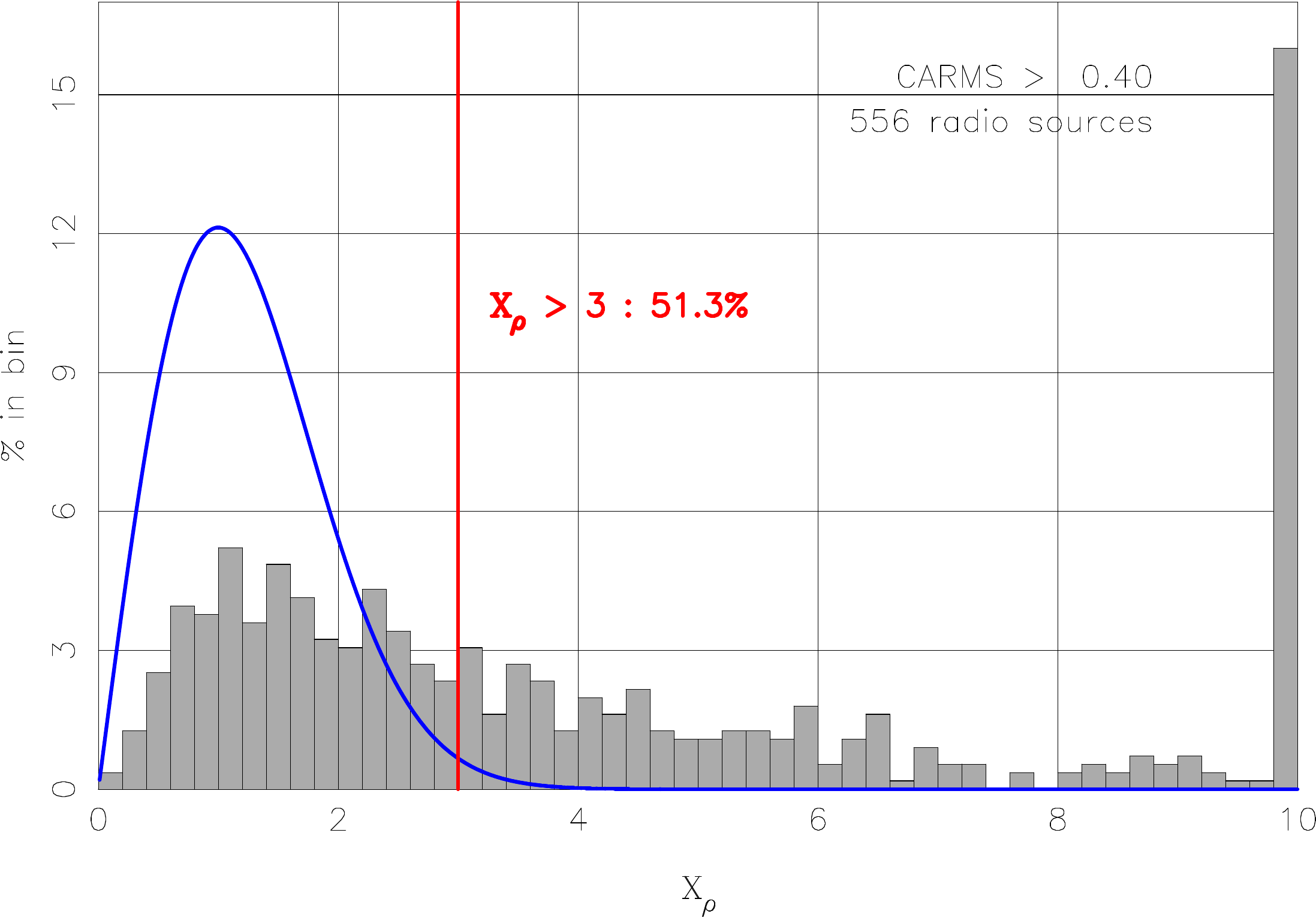}
   \caption{Histogram of $X_{\rho}$ for the 2490 sources (top), the sources with CARMS < 0.10 (middle), and
   the sources with CARMS > 0.40 (bottom).   
   The sources with $X_{\rho}$ > 10 are accounted in the last bin. The blue curves show the Rayleigh distributions with unit standard deviation. The total number of sources in each of the three samples is shown in black on the top-right of each panel, and the number of the sources with $X_{\rho}$ > 3.0 in red. The straight red lines correspond to $X_{\rho}$ = 3. The remarkable differences in the distributions of $X_{\rho}$ between these three groups of sources are the numbers of sources in the last bins, $X_{\rho}$ > 9.8.}
              \label{hist_normalized}%
    \end{figure*}

We examined the normalized arc lengths $X_{\rho}$ with respect to the CARMS values. 
The statistics of $X_{\rho}$ are shown in Table \ref{tab1_CARMS}.
A dependence of $X_{\rho}$ on the CARMS values is revealed.
Figure \ref{hist_normalized} shows the three distributions of $X_{\rho}$ for 
the 2460 common sources (top), the sources with CARMS < 0.10 (middle), and the sources with CARMS
> 0.40 (bottom). About 26 percent of the 2460 sources have $X_{\rho}$ > 3.
For the 207 radio sources with little structure, 
the distribution of $X_{\rho}$ shown in the middle panel is close to the expected Rayleigh distribution;
however, for the 556 radio sources with CARMS > 0.40, the distribution of $X_{\rho}$  
clearly deviates from the Rayleigh distribution --- half of the sources have $X_{\rho}$ 
> 3 and one sixth even have $X_{\rho}$ > 10. The probability of having statistically significant position differences 
is doubled for the radio sources with extended source structure (CARMS > 0.4). 

\begin{figure}[tbhp!]
   \resizebox{\hsize}{!}{\includegraphics{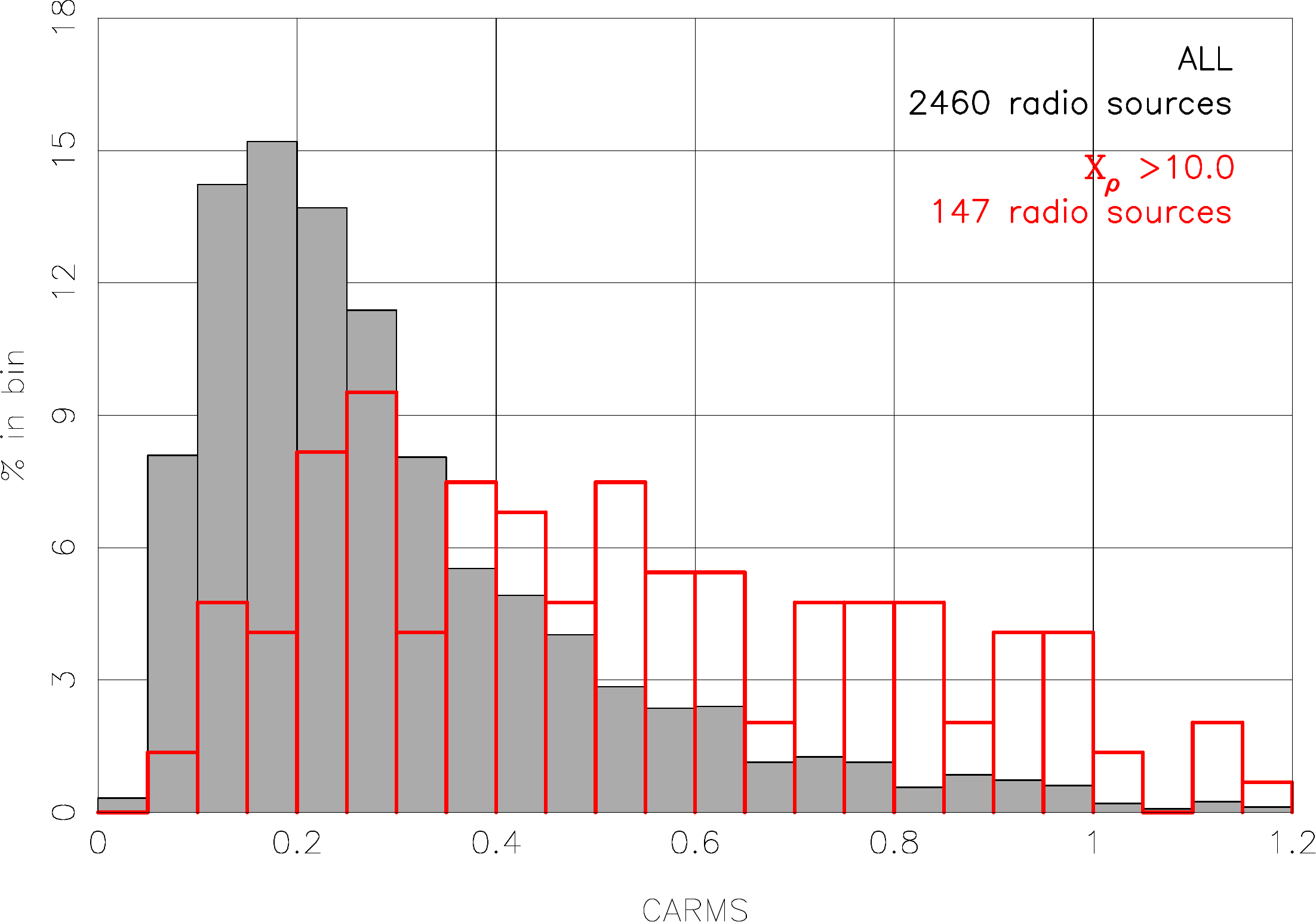}}
   \caption{Histogram of the CARMS values for the 2460 radio sources (filled gray bars) and for the 147 radio sources with $X_{\rho}$ > 10.0 (open red bars). About 3/5 of these 147 sources 
have the CARMS values larger than 0.40, while less than one quarter of the 2460 sources have the CARMS values larger than 0.40.}
              \label{histogram_04}%
    \end{figure}

Figure \ref{histogram_04} shows the distributions of the CARMS values for all 2460 sources (gray filled bins) and the 147 sources with $X_{\rho}$ > 10 (red open bins).  
The mean and median CARMS values for all 2460 sources are 
0.30 and 0.24, respectively; for the 147 sources with $X_{\rho}$ > 10 these values are 0.52 and 0.48. About 60 percent of the sources with $X_{\rho}$ > 10 have CARMS > 0.40. Given that only 23 percent of the 2460 sources have CARMS > 0.40, the high correlation is also identified between the sources with statistically significant position differences and the sources with extended structure. 

The differences of the CARMS values for the sources with various ranges of 
$\rho$ and $X_{\rho}$ are shown in Table \ref{arc_length_carms}. 
For different magnitudes of $\rho$, the mean and median CARMS values for the sources with $X_{\rho}$ > 4 
are all the largest among the three categories based on $X_{\rho}$; 
these values for the sources with $X_{\rho}$ in the range of 
3 to 4 are larger than for the sources with $X_{\rho}$ < 3. On average, the difference in the CARMS values is $\sim$0.2 between the sources with and without statistically significant position differences. There are only a slight increase in the mean and median CARMS values as $\rho$ increases for $X_{\rho}$ > 4. One should be cautious when interpreting the results in Table \ref{arc_length_carms}, because they will change with better uncertainties of source positions available in future $Gaia$ data releases. With the significant improvement in position
uncertainties expected from the $Gaia$ observations, the sources with current $X_{\rho} \leq$ 3 
can have $X_{\rho} >$ 3, as happened for the $Gaia$ DR2 compared
to the $Gaia$ DR1 and for $Gaia$ EDR3 compared to the $Gaia$ DR2. Meanwhile, the arc lengths $\rho$ for the 1813 sources with $X_{\rho} \leq$ 3 will generally decrease, which can be demonstrated by the $Gaia$ DR2 and EDR3. The arc lengths of these 1813 sources are all smaller than 4.0\,mas; the number of the sources with $\rho$ $\geq$ 4.0\,mas should 
not be changed dramatically, unless a few new
common sources between $Gaia$ and the ICRF3 will be identified from the future $Gaia$ data releases. Since the ICRF3 sources were systematically included in the $Gaia$ quasar list, those missing sources in the $Gaia$ EDR3 are probably too faint in optical and it is unlikely to have significantly more matches from $Gaia$.
As shown in Table \ref{tab1_CARMS}, the uncertainties of $\rho$ have mean and median values of about 0.3\,mas
and 0.2\,mas, which allow the large $\rho$, for instance larger than 4.7\,mas, 
be confidently detected but are not able to fully identify the sources with
$\rho$ < 1.0\,mas. Therefore, when the final $Gaia$ data release is available 
to identify more sources with small $\rho$ and large
$X_{\rho}$, the mean and median CARMS values will thus decrease for the 
sources with $\rho$ < 1.0\,mas and $X_{\rho}$ > 4. 
We would expect to have the CARMS values steady increasing with respect to $\rho$ in the future $Gaia$ data releases, as we see that $\rho$ increases with CARMS in Table~\ref{tab1_CARMS}. In the following investigation, we set the limit of $X_{\rho}$ = 4.0, at the 99.994\,$\%$ confidence level, to identify the sources with 
statistically significant position differences. 

%

\begin{table*}[tbhp!]
\caption{CARMS values with respect to $\rho$ and $X_{\rho}$}
\label{arc_length_carms}
\centering
\begin{tabular}{crcccrcccrcc}
\hline\noalign{\smallskip}
\multirow{2}{*}{$\rho$ [mas]} & \multicolumn{3}{c}{if ( $X_{\rho}$ > 4 )}& &\multicolumn{3}{c}{if ( 4 $\geq$ $X_{\rho}$ > 3 )}& &\multicolumn{3}{c}{if ( $X_{\rho}$ $\leq$ 3 )}\\\noalign{\smallskip}\cline{2-4}\cline{6-8}\cline{10-12}\noalign{\smallskip}
&$N_{\texttt{src}}$& Mean & Median&&$N_{\texttt{src}}$& Mean & Median&&$N_{\texttt{src}}$& Mean & Median\\
\noalign{\smallskip}\hline\noalign{\smallskip}
  < 0.4                &  34  &  0.43  &  0.44 & & 33 & 0.35 & 0.30 && 1011 &0.26 & 0.23 \\
$[$  0.4 --\ \ \ 0.7 )  &  56 &   0.43   & 0.36 & &62 & 0.33 &0.28&& 425 & 0.25 & 0.21\\
$[$  0.7 --\ \ \ 1.0 ) &  55  & 0.47 &   0.45 & &48 & 0.25 & 0.21 && 179 & 0.27 & 0.21\\
$[$  1.0 --\ \ \ 2.0 )  & 114  &  0.50  &  0.43 && 52 &0.30 &0.19 && 164 & 0.25 & 0.20\\
$[$  2.0 --\ \ \ 4.0 )  & 98 &  0.40  &  0.35 && 16 & 0.50 & 0.43 && 34 & 0.26 & 0.20\\
$[$  4.0 --\ \ \ 7.0 ) &  44   & 0.46   & 0.44 & & 4 & 0.36 & 0.33 & & 0 & $\ldots$  & $\ldots$ \\
$\geq$ 7.0 &    31 &   0.51  &  0.47 & & 0 & $\ldots$ & $\ldots$ & & 0 & $\ldots$ & $\ldots$ \\
\noalign{\smallskip}\hline\noalign{\smallskip}
all & 432 & 0.46 & 0.42 & & 215 & 0.32 & 0.27 && 1813 & 0.26 & 0.22\\
\noalign{\smallskip}\hline
\end{tabular}
\end{table*}

%

\subsection{Optical $G$ magnitude and redshift}
We examined optical $G$ magnitude and redshift $z$ with respect to the 
CARMS values to investigate if there is any potential correlation between 
the CARMS values and the optical properties. Table \ref{magnitude_carms}
shows the statistics of $G$ and $z$.
Both the mean and median magnitudes generally decrease with respect to 
the CARMS values; the difference in $G$ between
the sources with CARMS < 0.1 and with CARMS > 0.9 is about 0.9\,mag.
Based on the high correlation between the radio luminosity and the optical
luminosity is shown in \cite{2010A&A...520A..62A},
the sources with higher luminosity at optical wavelengths will have 
higher radio flux densities. 
One can also expect a correlation between radio luminosity and extended structure 
which is driven by jet power --- larger power means higher radio luminosity and more 
extended structure in linear scale due to the jet being able to drill its path.
Radio sources with higher flux densities tend to have more extended structure and consequently 
larger CARMS values, as shown for
the 30 most frequently observed sources in geodetic VLBI by \cite{xu2019apjs}. 
Since the CRF sources are flux-limited, at large redshifts the sources must have  
high luminosity, and consequently their powers and extents are larger 
than at low redshift. This should partly explain the correlation between $z$ and CARMS in the table. 
The correlation between CARMS and both 
$G$ and $z$ seems to be significant. 

\begin{table*}[tbhp!]
\caption{Optical $G$ magnitude and redshift $z$}
\label{magnitude_carms}
\centering
\begin{tabular}{crcccrcc}
\hline\noalign{\smallskip}
\multirow{2}{*}{CARMS} & \multicolumn{3}{c}{Optical $G$ magnitude [mag]}& &\multicolumn{3}{c}{Redshift $z$}\\\noalign{\smallskip}\cline{2-4}\cline{6-8}\noalign{\smallskip}
&$N_{\texttt{src}}$& Mean & Median&&$N_{z}$& Mean & Median\\
\noalign{\smallskip}\hline\noalign{\smallskip}
  < 0.1            &  207 & 19.283 & 19.486 &&  181 &  1.270 &  1.062\\
$[$  0.1 --  0.2 ) &  724 & 18.965 & 19.112 &&  624 &  1.188 &  1.072\\
$[$  0.2 --  0.3 ) &  617 & 18.665 & 18.759 &&  553 &  1.214 &  1.139\\
$[$  0.3 --  0.4 ) &  334 & 18.712 & 18.841 &&  305 &  1.355 &  1.292\\
$[$  0.4 --  0.5 ) &  220 & 18.460 & 18.498 &&  204 &  1.350 &  1.256\\
$[$  0.5 --  0.6 ) &  128 & 18.489 & 18.581 &&  116 &  1.413 &  1.312\\
$[$  0.6 --  0.7 ) &   87 & 18.381 & 18.474 &&   78 &  1.268 &  1.203\\
$[$  0.7 --  0.8 ) &   59 & 18.442 & 18.446 &&   53 &  1.622 &  1.400\\
$[$  0.8 --  0.9 ) &   35 & 18.678 & 18.740 &&   33 &  1.506 &  1.460\\
$\geq$  0.9        &   49 & 18.359 & 18.597 &&   47 &  1.434 &  1.351\\
 \noalign{\smallskip}\hline\noalign{\smallskip}
  all & 2460 & 18.763 & 18.910 && 2198 &  1.275 &  1.182\\
\noalign{\smallskip}\hline
\end{tabular}
\end{table*}

We further examined $G$ and $z$ in more detail.
This investigation can be biased, because the uncertainties of $Gaia$ positions 
depend on $G$, as shown in \cite{2018A&A...616A..14G}.
The statistics of arc lengths and normalized arc lengths with respect to $G$ can be dramatically changed when 
new position estimates with improved uncertainties are available from $Gaia$ in the near future.
We nevertheless attempt to address it based on the $Gaia$ EDR3.

Table \ref{magnitude_1960} shows the statistics of arc lengths,
the major axes of the error ellipses of the $Gaia$ positions and the VLBI positions, the CARMS values, and $z$ with respect to different optical $G$ magnitudes for the 2028 sources with $X_{\rho}$ $\leq$ 4.
As we expect, the $G$ and the $z$ are positively
correlated for these sources --- when object is further away, it appears dimmer. 
The differences of the mean CARMS values at various 
ranges of $G$ are no larger than 0.06 and those of the median values are no larger than 
0.08. There is a small decrease in the CARMS values when $G$ increases, which demonstrates that
when a source locates farther away the scale of its structure may decrease. 
The magnitudes of $\rho$ gradually  
increase with respect to $G$, however, the position uncertainties of both $Gaia$ and VLBI also vastly increase.
Since the ratio of the arc lengths to its uncertainties is always at the same level for different ranges of $G$,
it is not possible from the result to conclude that there is dependence of $\rho$ on $G$.

\begin{table*}[tbhp!]
\caption{Statistics of the 2028 sources with $X_{\rho} \leq$ 4.}
\label{magnitude_1960}
\centering
\begin{tabular}{crcccrcccccrcc}
\hline\noalign{\smallskip}
\multirow{2}{*}{$G$ [mag]} & \multirow{2}{*}{$N_{\texttt{src}}$} & \multicolumn{2}{c}{$\rho$ [mas]} && \multicolumn{2}{c}{$\sigma_{\texttt{pos,max}}$ [mas]}&&\multicolumn{2}{c}{CARMS}&& \multicolumn{3}{c}{$z$}
\\\noalign{\smallskip}\cline{3-4}\cline{6-7}\cline{9-10}\cline{12-14}\noalign{\smallskip}
&& Mean & Median&& $Gaia$ & VLBI&&Mean & Median&&$N_{z}$ & Mean & Median\\
\hline\noalign{\smallskip}
  < 15.0           &    7 &  0.304 &  0.256 && 0.020   & 0.218 &&    0.25  &  0.21&&    7 &  0.304 &  0.200 \\
$[$ 15.0 -- 16.0 ) &   35 &  0.258 &  0.175 && 0.030   & 0.175 &&    0.30  &  0.27&&   35 &  0.459 &  0.310 \\
$[$ 16.0 -- 16.5 ) &   29 &  0.316 &  0.244 && 0.043   & 0.199 &&    0.29  &  0.26&&   29 &  0.711 &  0.557 \\
$[$ 16.5 -- 17.0 ) &   62 &  0.283 &  0.218 && 0.051   & 0.177 &&    0.28  &  0.24&&   59 &  1.014 &  1.003 \\
$[$ 17.0 -- 17.5 ) &  125 &  0.337 &  0.271 && 0.072   & 0.199 &&    0.29  &  0.25&&  120 &  1.029 &  0.954 \\
$[$ 17.5 -- 18.0 ) &  176 &  0.312 &  0.215 && 0.094   & 0.186 &&    0.29  &  0.25&&  172 &  1.211 &  1.093 \\
$[$ 18.0 -- 18.5 ) &  265 &  0.347 &  0.288 && 0.126   & 0.210 &&    0.28  &  0.24&&  251 &  1.261 &  1.200 \\
$[$ 18.5 -- 19.0 ) &  312 &  0.392 &  0.331 && 0.179   & 0.207 &&    0.27  &  0.23&&  292 &  1.452 &  1.384 \\
$[$ 19.0 -- 19.5 ) &  331 &  0.492 &  0.399 && 0.247   & 0.230 &&    0.25  &  0.21&&  301 &  1.423 &  1.375 \\
$[$ 19.5 -- 20.0 ) &  315 &  0.611 &  0.491 && 0.370   & 0.223 &&    0.24  &  0.19&&  259 &  1.428 &  1.300 \\
$[$ 20.0 -- 20.5 ) &  275 &  0.917 &  0.790 && 0.623   & 0.228 &&    0.24  &  0.19&&  202 &  1.502 &  1.375 \\
$\geq$        20.5 &   96 &  1.687 &  1.434 && 1.079   & 0.272 &&    0.27  &  0.20&&   64 &  1.274 &  0.980 \\
 \noalign{\smallskip}\hline\noalign{\smallskip}
 all & 2028 &  0.633 &  0.454 && 0.293   & 0.216 &&    0.26  &  0.22&& 1791 &  1.314 &  1.219 \\
\noalign{\smallskip}\hline
\end{tabular}
\begin{tablenotes}
\item \textit{Note}. The values in the fifth and sixth columns are the mean $\sigma_{\texttt{pos,max}}$ for $Gaia$ and VLBI position estimates, respectively. 
 \end{tablenotes}
\end{table*}

Table \ref{magnitude_389} shows the statistics of the same quantities as Table~\ref{magnitude_1960} but 
for the 432 sources with $X_{\rho}$ > 4. The arc lengths increase by a factor 
of $\sim$10 from $G$ < 15 mag to $G$ $\geq$ 20 mag. This apparent dependence
of $\rho$ on $G$, however, is mainly due to the high correlation between the $Gaia$ position 
uncertainties and $G$, as shown in the Table. Because the $Gaia$ position uncertainties
get worse dramatically as $G$ becomes higher, a uniformed threshold of $X_{\rho}$, which is 4 in the study, forces only 
the sources with large enough arc lengths to be selected at the higher optical magnitudes. As discussed before,
these statistics will be changed with new position estimates available from the future $Gaia$ data releases. 
%

%
\begin{table*}[tbhp!]
\caption{Statistics of the 432 sources with $X_{\rho}$ > 4.}
\label{magnitude_389}
\centering
\begin{tabular}{crcccrcccccrcc}
\hline\noalign{\smallskip}
\multirow{2}{*}{$G$ [mag]} & \multirow{2}{*}{$N_{\texttt{src}}$} & \multicolumn{2}{c}{$\rho$ [mas]} && \multicolumn{2}{c}{$\sigma_{\texttt{pos,max}}$ [mas]}&&\multicolumn{2}{c}{CARMS}&& \multicolumn{3}{c}{$z$}
\\\noalign{\smallskip}\cline{3-4}\cline{6-7}\cline{9-10}\cline{12-14}\noalign{\smallskip}
&& Mean & Median&&Gaia & VLBI&&Mean & Median&& $N_{z}$ & Mean & Median\\
\hline\noalign{\smallskip}
  < 15.0           &   9 &   0.369 &  0.269 &&  0.016 & 0.053 &&   0.42 &  0.30 &&  9 &  0.228 &  0.160 \\
$[$ 15.0 -- 16.0 ) &   17 &  0.978 &  0.521 &&  0.032 &  0.074 &&   0.59 &  0.63 &&  17 &  0.394 &  0.302 \\
$[$ 16.0 -- 16.5 ) &   19 &  1.083 &  0.839 &&  0.040 &  0.121 &&   0.56 &  0.60 &&  18 &  1.182 &  1.258 \\
$[$ 16.5 -- 17.0 ) &   34 &  3.820 &  0.867 &&  0.062 &  0.155 &&   0.47 &  0.43 &&  34 &  1.332 &  1.140 \\
$[$ 17.0 -- 17.5 ) &   43 &  1.810 &  0.868 &&   0.088 &  0.125 &&   0.48 &  0.43 &&  43 &  1.093 &  0.994 \\
$[$ 17.5 -- 18.0 ) &   59 &  1.581 &  0.951 &&   0.098 &  0.157 &&   0.47 &  0.44 &&  57 &  1.283 &  1.285 \\
$[$ 18.0 -- 18.5 ) &   68 &  2.483 &  1.505 &&  0.146 &  0.167 &&   0.44 &  0.39 &&  68 &  1.228 &  1.208 \\
$[$ 18.5 -- 19.0 ) &   52 &  4.258 &  1.535 &&  0.184 &  0.229 &&   0.44 &  0.41 &&  48 &  1.065 &  0.949 \\
$[$ 19.0 -- 19.5 ) &   53 &  4.777 &  2.289 &&  0.250 &  0.222 &&   0.46 &  0.41 &&  44 &  1.061 &  0.726 \\
$[$ 19.5 -- 20.0 ) &   35 &  4.953 &  2.987 &&   0.405 &  0.241 &&   0.44 &  0.35 &&  31 &  1.054 &  0.667 \\
$[$ 20.0 -- 20.5 ) &   28 &  5.344 &  3.623 &&  0.622 &  0.255 &&   0.31 &  0.30 &&  26 &  0.940 &  0.770 \\
$\geq$ 20.5        &   15 &  4.176 &  3.151 &&   0.805 &  0.292 &&   0.49 &  0.46 &&   12 &  1.254 &  1.037 \\
 \noalign{\smallskip}\hline\noalign{\smallskip}
  all              &  432 &  3.173 &  1.510 &&  0.207 &  0.183 &&   0.46 &  0.42 && 407 &  1.103 &  0.901 \\
\noalign{\smallskip}\hline
\end{tabular}
\begin{tablenotes}
\item \textit{Note}. The values in the fifth and sixth columns are the mean $\sigma_{\texttt{pos,max}}$ for $Gaia$ and VLBI position estimates, respectively. 
 \end{tablenotes}
\end{table*}

\begin{figure}
   \resizebox{\hsize}{!}{\includegraphics{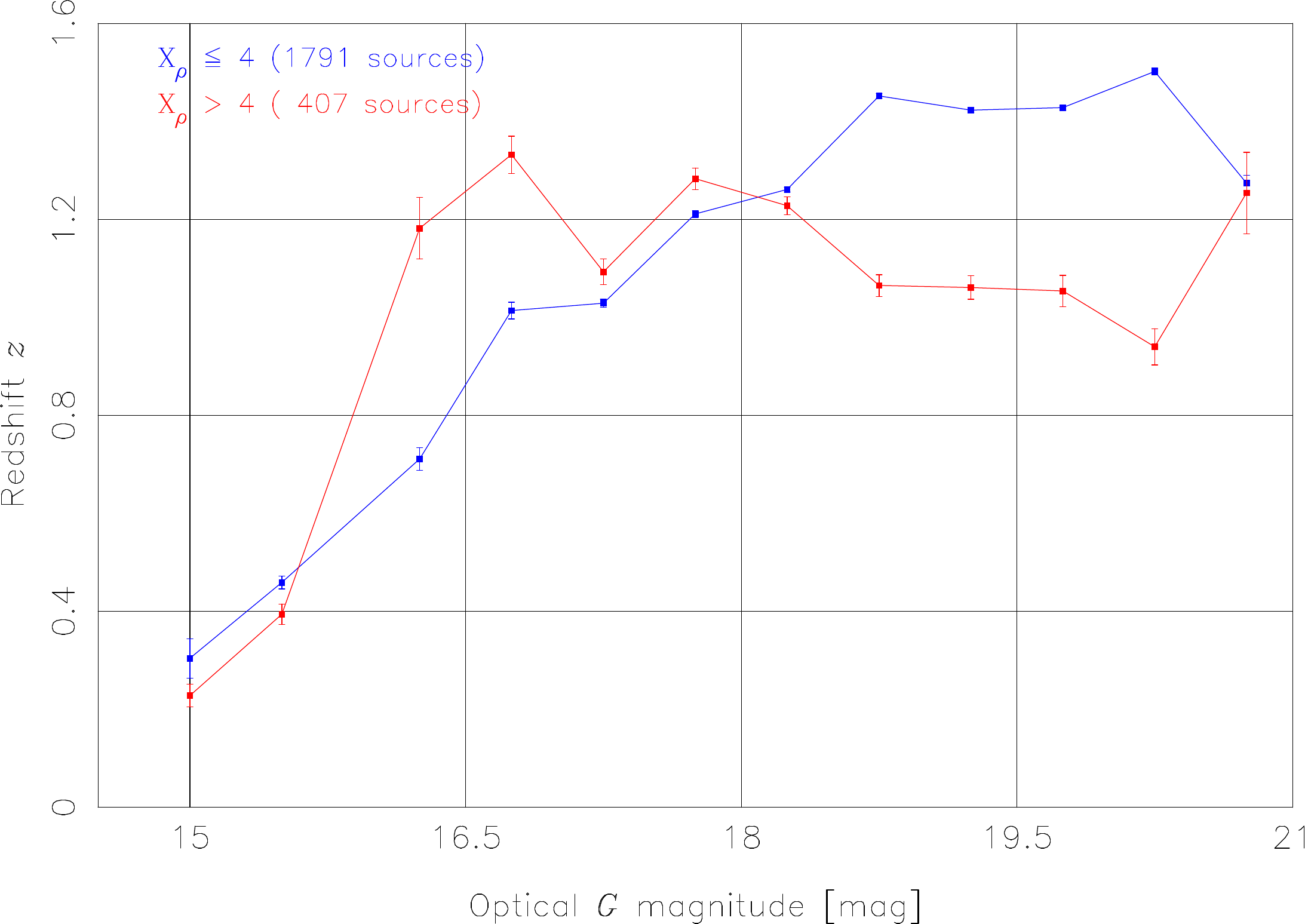}}
   \caption{Mean redshift values with respect to the optical $G$ magnitudes for the 2198 sources with their redshifts available. The blue curve is for the sources with $X_{\rho}$ $\leq$ 4, and the red curve is for the sources with $X_{\rho}$ > 4. The error bars show the estimated uncertainties of the mean values. 
The bin windows of $G$ are shown in the first column in Table \ref{magnitude_1960}. 
The sources with $X_{\rho}$ > 4 have substantially lower $Z$ at $G$ > 18.5\,mag but higher $z$ at $G\simeq16.5$\,mag
than the sources with $X_{\rho} \leq$ 4. The statistics are shown in Tables \ref{magnitude_1960} and \ref{magnitude_389}.}
              \label{G_redshift}%
    \end{figure}

By comparing the results in Tables \ref{magnitude_1960} and \ref{magnitude_389}, the major differences of these
two groups of sources are found to be CARMS and $z$. The CARMS values of the sources
with $X_{\rho}$ > 4 are larger by $\sim$0.2 than those of the sources with $X_{\rho}$ $\leq$ 4; 
the mean and median $z$ values are smaller by 0.21 and 0.32, respectively. The relationship between 
$G$ and $z$ for these two groups of sources are shown in Fig. \ref{G_redshift}.
The sources with $X_{\rho} \leq$ 4 have the $z$ steady increasing 
over $G$, while the sources with $X_{\rho}$ > 4 even have a small decrease in $z$ when $G$ > 16.5\,mag. 

We argue that the statistically significant position
differences may also be associated with, for instance, some weak but nearby (small $z$) optical objects.

\section{Discussion}
\subsection{Radio source structure}
The CRF sources have radio emission with angular scales at mas levels over the sky, called source structure. 
It causes structure delays up to
hundreds of picoseconds as shown in modeling by \cite{1990AJ.....99.1309C} and in actual observations by \cite{2016AJ....152..151X}. Based on the CONT14 observations\footnote{\url{https://ivscc.gsfc.nasa.gov/program/cont14/}}, 
\cite{2018JGRB..12310162A} suggested that source structure $is$ the major contributor to errors in the astrometric/geodetic VLBI.
Since these effects in VLBI group delays have not been modeled in the VLBI data analysis, based on which the ICRFs were built and maintained, the source positions from VLBI change over time due to both the different observing geometry between antennas and sources and the varying structure. For a large fraction of CRF sources, the structure effects can change their positions at the level of 0.5\,mas, as shown in the position time series of 39 well-observed sources \citep[][see the plots in the IERS Technical Note 35\footnote{\url{https://www.iers.org/SharedDocs/Publikationen/EN/IERS/Publications/tn/TechnNote35/tn35_017.pdf?__blob=publicationFile&v=1}}]{2009ITN....35....1M}. The number of sources affected by the structure effects will dramatically increase when we consider the position differences between $Gaia$ and VLBI down to the levels of $\sim$0.3\,mas. Based on the CARMS values, 40 percent of CRF sources have significant structure. 

CARMS tells the structure effects in amplitude observables. 
For a source with CARMS = 0.1, the ratios of the amplitude observables over various 
combinations of quadrangle have an rms of 1.1. Those ratios have an rms of 1.5 for CARMS = 0.4, 
and 1.8 for CARMS = 0.6. It is straightforward to understand that the source with a small
CARMS value is close to point-like, and with a large CARMS value has extended structure. 

\begin{figure*}
   \centering
        \includegraphics[width=0.49\textwidth]{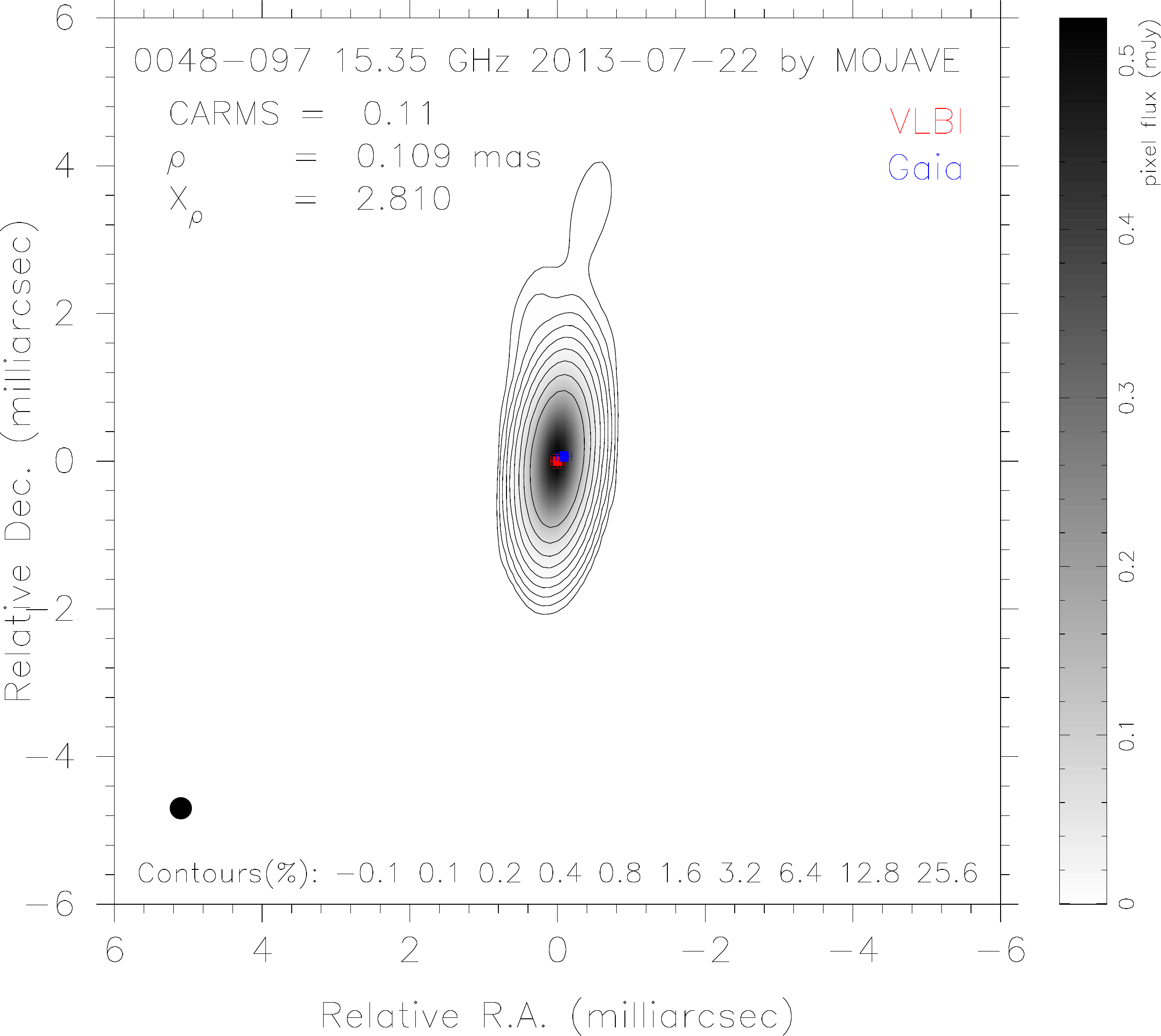}
   \includegraphics[width=0.49\textwidth]{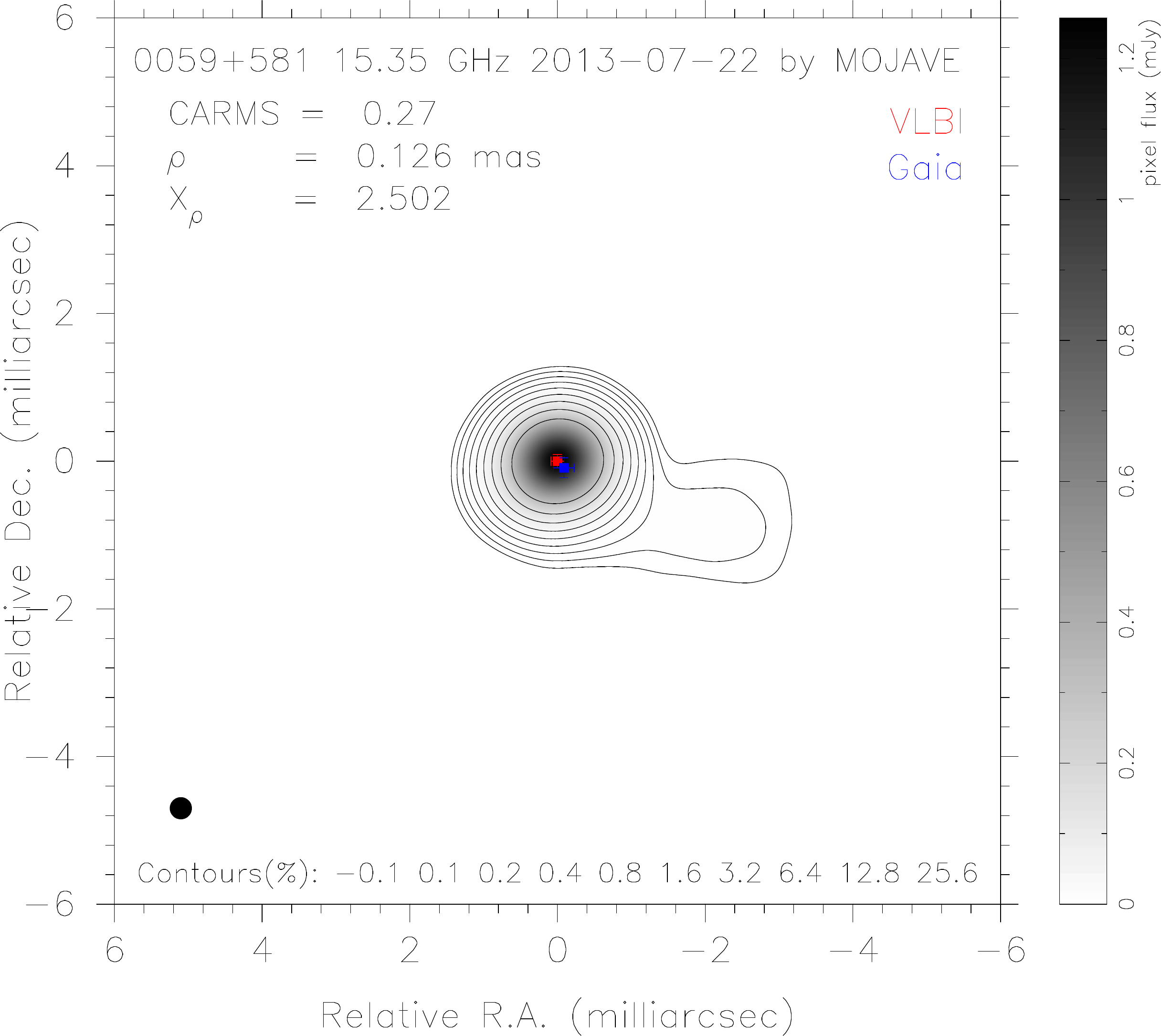}
                \includegraphics[width=0.49\textwidth]{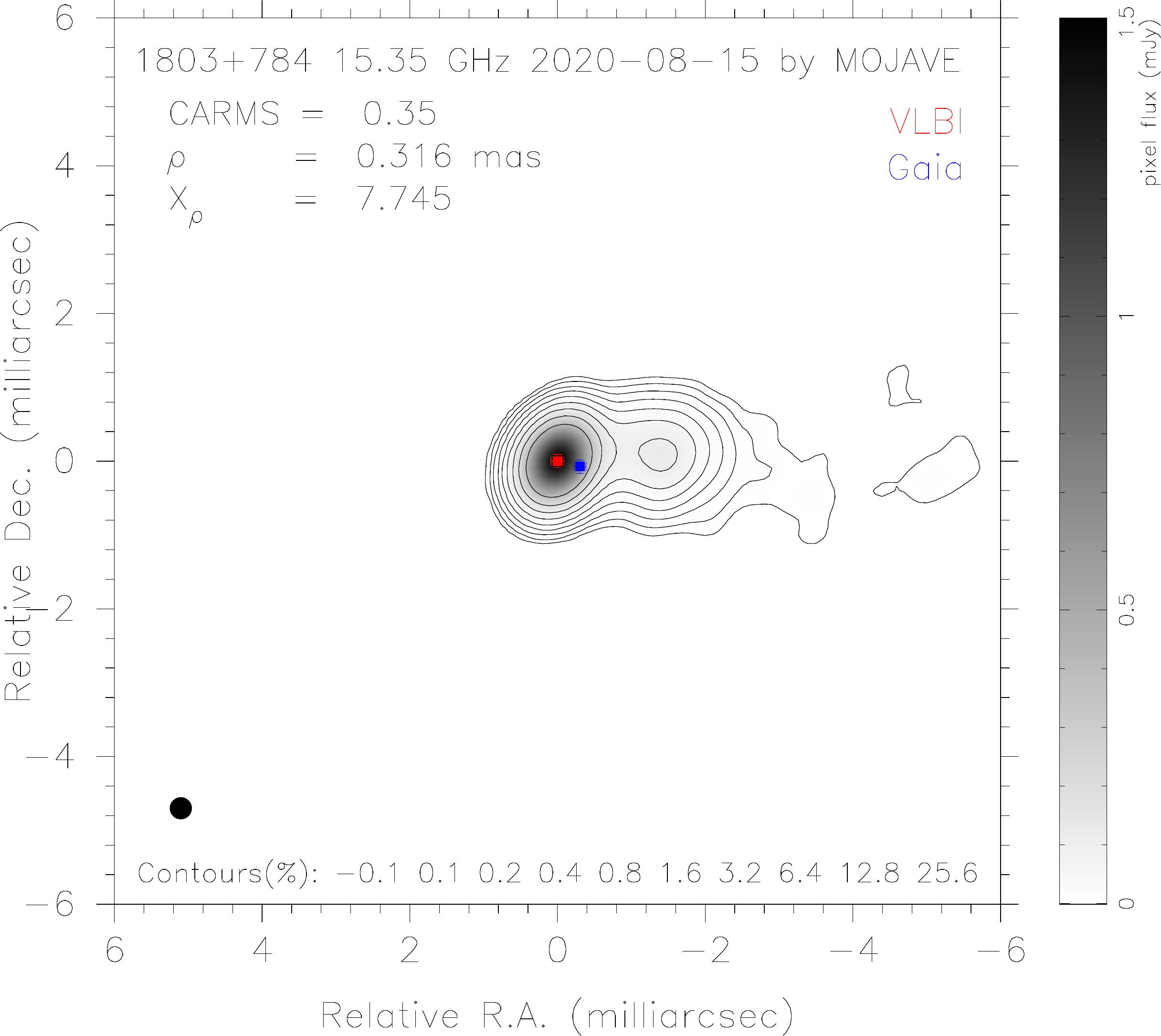}
                    \includegraphics[width=0.49\textwidth]{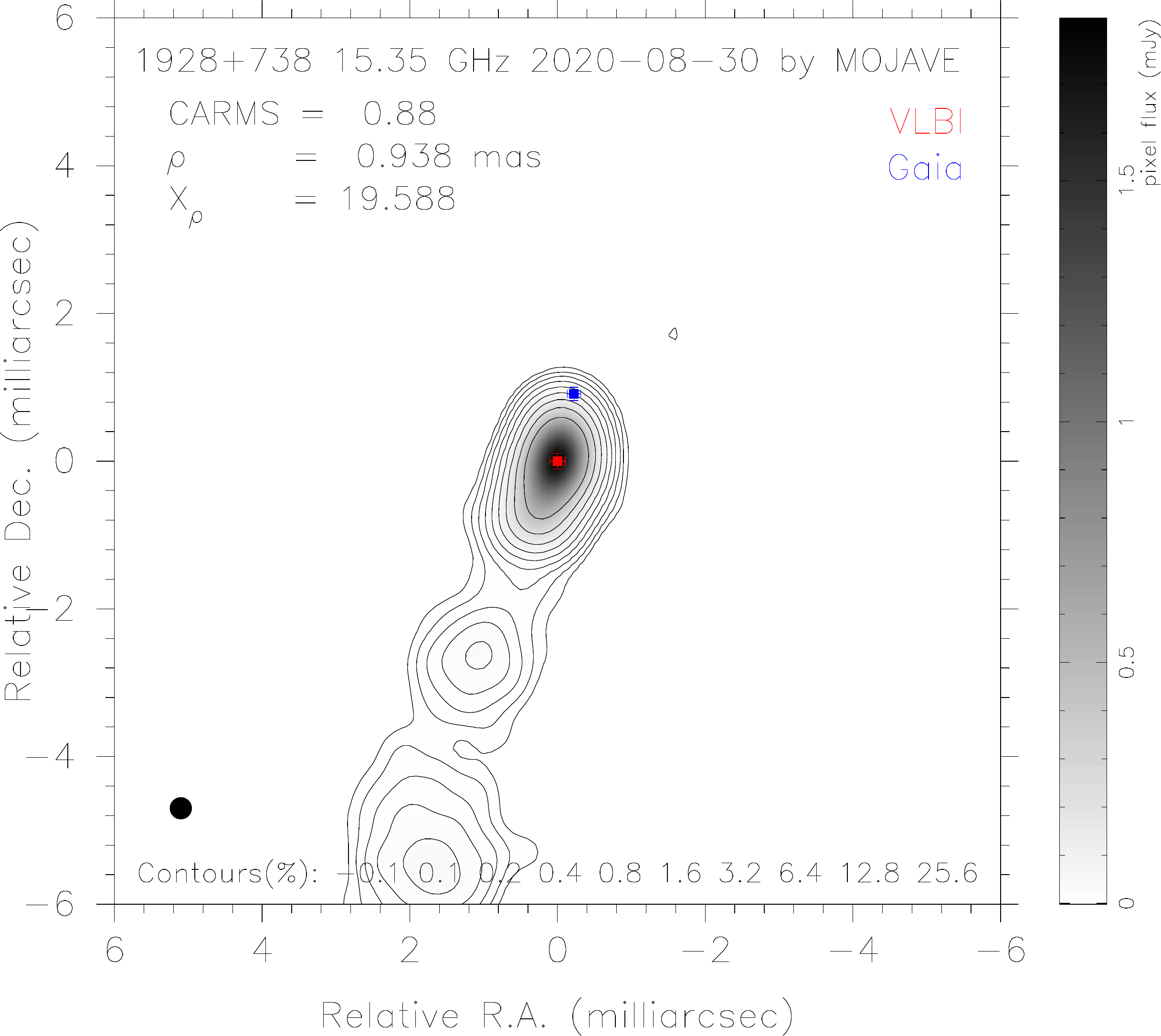}
   \caption{MOJAVE images of four sources, 0048$-$097 (CARMS=0.11, upper-left), 0059+581 (CARMS=0.27, upper-right),
   1803+784 (CARMS=0.35, bottom-left), and 1928+738 (CARMS=0.88, bottom-right). These images were made based on VLBA observations at 15.35 GHz by the MOJAVE project. The peaks of flux are selected as the origins. The VLBI positions are formally assumed to be at the origins as shown by the red dots. Since the MOJAVE images and the ICRF3 are derived from observations at different frequencies , this assumption may introduce systematic errors. Based on their position differences between the $Gaia$ EDR3 and the ICRF3, the $Gaia$ positions are thus located at the blue dots. The error bars are the 3$\sigma$ uncertainties of right ascension and declination from $Gaia$ and VLBI. It is also conspicuous that when $X_{\rho}$ > 4 the VLBI to $Gaia$ position vectors favor the directions along and opposite to the jets, as shown by \citep{2017A&A...598L...1K,2019MNRAS.482.3023P}. The $\rho$ and $X_{\rho}$ values are shown on the upper-left corner of each plot. These four plots demonstrate how the scales of the structure look like in terms of different CARMS values. Nevertheless, we should mention that the CARMS values and the ICRF3 are based on VLBI observations at the frequency band around 8.4 GHz over 40 years, while these images were made from observations at 15.35 GHz during the short periods shown at the top of each plot. The jet components
always become more prominent at the lower frequency bands. The images were convolved with a circular beam of 0.3 mas as indicated by the black circle in the bottom left corner, about 40$\%$ of the typical MOJAVE beam size. Overlay contours are shown 
at ten levels of peak percentage specified in the bottom of plots.}
              \label{mojave}%
    \end{figure*}

In Fig. \ref{mojave}, we show the images from MOJAVE for four sources, 0048$-$097, 0059$+$581, 1803$+$784, and 1928$+$738. Since the VLBI observations for deriving the images are at different frequencies by different antenna arrays during different time periods compared to the observations for the ICRF3 and the CARMS values, we cannot expect an exact proportional relation between the CARMS values and the scales of the MOJAVE images. However, they are already of great help to demonstrate the differences between the CARMS values smaller and larger than 0.3.
The two sources 0048$-$097 (CARMS=0.11) and 0059$+$581 (CARMS=0.27) have virtually compact cores only,
whereas the other two sources, 1803$+$784 (CARMS=0.35) and 1928$+$738 (CARMS=0.88), have significant emissions
from the jets at mas scales.
The relative positions between $Gaia$ and VLBI are also shown in the plots. It is obvious in the plots that the $Gaia$-VLBI position differences are typically parallel to the jet directions, which has already been reported by \cite{2017A&A...598L...1K} and \cite{2019MNRAS.482.3023P} and will be discussed in Sec. \ref{directions}.
 
There are four remarks concerning CARMS. First, it was
calculated based on actual VLBI observations
rather than based on the maps of radio sources. 
Once the CARMS is large, the source should have extended structure;
but if the source has extended structure, it does not necessarily have a large CARMS value due to insufficient
observations in terms of $(u,v)$ coverage 
to capture the structure. However, the great advantage of using actual VLBI observations 
is that it quantifies the magnitude of structure effects over the whole 
time period of 40 years. Second, CARMS is based on (log) closure amplitudes, 
which are not sensitive to the absolute source position. 
Therefore, only the relative structure, i.e., the relative positions
and the relative fluxes between the multiple components, is defined by CARMS; if a source with compact structure
changes its position on the sky, the CARMS value cannot tell that change. Third, since there was no
attempt to do proper weighting for different sizes of quadrangle and select an independent set of closure amplitudes for each individual source in deriving CARMS values, it becomes difficult 
to tell a source with a medium CARMS value, 0.25--0.30, as having structure to what extent. Fourth, CARMS was derived from the X-band observations only, while the ICRFs are based on the ionosphere-free delays through the linear combination of the group delays at the S/X band. The structure effects in the S-band observations 
thus are ignored in this study. Even though the contribution of the structure effects at the S-band is scaled down by a factor of $\sim$13.8 in that  linear combination process, it can be significant for some radio sources.
These should partly explain why there are sources with CARMS < 0.10 but with $X_{\rho}$ > 3.0, as shown in
the middle panel of Fig. \ref{hist_normalized}.

Modeling structure effects is still missing in astrometric/geodetic VLBI data analysis after it has been discussed for several decades.
The practical problems are to continuously make images for hundreds of sources and for each source many times if structure changes. 
The main challenge is that the images for modeling structure effects have to be registered over time for each source in order to maintain a stable CRF at high accuracy. The next generation of geodetic VLBI, known as VGOS \citep{niell2007,petrachenko2009}, requires to register the images of each source at the four different bands in the range of 3.0--14.0\,GHz \citep{10.1002/essoar.10504599.1}.
Otherwise, only the relative structure effects can be reduced, and the misalignment of the images at different epochs or at different frequency bands due to core shift, discussed in the next section, inevitably leads to source position variations. Due to the limitation in
imaging resolutions, identifying the reference points in 
structure/images is difficult for the accuracy levels better than 0.1\,mas.
Therefore, aligning the images and investigating core shift are very crucial in order to mitigate these systematic effects.

\subsection{Core shift}

Source structure is frequency-dependent due to two factors: (1) the steep spectrum of the extended jet causing
the sources to have larger scales at lower frequencies; and (2) synchrotron self-absorption causing changes in the
optical depth along the jet. The latter factor leads to changes in the position of the core, where the optical depth is unity, 
depending on the observing frequency. This effect, so called core shift, was predicted by~\citet{1979ApJ...232...34B}. When the observing frequency increases, 
it causes the position of the core to move towards the jet base. 

Core shift was first measured for the source 1038$+$528A with a magnitude 
of $\sim$0.7\,mas at 2.3\,GHz and 8.4\,GHz by referring to its 
nearby source 1038$+$528B \citep{1985A&A...142...71M}. Since then, it has been 
measured for 29 sources with a median value of 0.44\,mas between 2.3\,GHz and 8.4\,GHz by \cite{2008A&A...483..759K}, 20 sources with a median value of 1.21 mas between 1.4\,.GHz and 15.4\,GHz and 0.24\,mas 
between 5.0\,GHz and 15.4\,GHz by \cite{2011A&A...532A..38S}, 163 sources with a median value of
0.128\,mas between 8.4\,GHz and 15\,GHz \cite{2012A&A...545A.113P} and 40 sources with a typical value of 0.5\,mas between 2.3\,GHz and 8.4\,GHz \cite{2019MNRAS.485.1822P}. The frequency-dependency of the core position can be parameterized by $k\nu^{-\beta}$, where $k$ is a source-dependence core shift parameter --- it can be variable over time according to the study of \cite{2019MNRAS.485.1822P}, $\nu$ is the observing frequency, and $\beta$ is an astrophysical parameter. So far, $\beta$ is measured to be close to 1 
\citep{1998A&A...330...79L,2011A&A...532A..38S}, which agrees with the prediction under the condition of the 
equipartition between jet particle and magnetic field energy densities \citep{1979ApJ...232...34B}. 

The impact of core shift on astrometric positions measured by VLBI 
was discussed by \citet{porcas2009}, using a simple model of a point-source core.
Based on the median core shift between 2.3\,GHz and 8.4\,GHz, 0.44\,mas, from \cite{2008A&A...483..759K},
the core position is shifted by 0.166\,mas at the frequency of 8.4\,GHz and 
varies by 0.014\,mas over the frequency band of 8.2--8.9\,GHz used in 
most of geodetic VLBI observations. The position shift of 0.166\,mas can 
cause visibility phase variations of several degrees over the band, 
which are canceled out exactly by 
the additional phase variations due to the position shifts of 0.014\,mas over the band.
It was shown that given $\beta = 1$, group delays of observations on a point-like source refer to 
a fixed point at the jet base at any frequency and at any time, no matter whether 
$k$ varies or not over time. It is therefore believed that core shift
will not contribute to the position differences between $Gaia$ and VLBI, given that
$\beta \simeq 1$. 

Our special concern about core shift is not only the robust validation of $\beta \simeq 1$ for the CRF sources,
but also the simple source model used in \citet{porcas2009}. Core shift has two 
effects on source structure: (1) moving the absolute position of the core towards
the jet base when the frequency increases; and (2) changing the relative positions between the core and 
the jet components in structure. Apparently, the discussion of \citet{porcas2009}
investigated the first effect only. The truth is again that almost all the CRF sources have structure
at the mas scales, which changes over time. In the previous discussion, 
the relative positions between the
core and the jets will also be changed by amount 
of 0.014\,mas over the band to the opposite direction
of the absolute position shift of the core. 
The cancellation of the across-band phase variations in the point-source case breaks down for extended sources. 
Therefore, core shift can influence the position estimates 
determined from VLBI group delays. In this context, even though there may 
be no real connection between the magnitude of core shift and the scales of source structure, the impact of core shift will correlate with structure effects --- extended sources with large source structure effects tend to have larger core shift effect in the position differences between radio and optical 
than the sources with minimum structure.
Further studies are needed to verify this assumption.

\subsection{Sources with $\rho$ > 4.0\,mas and $X_{\rho}$ > 4}

There are 75 sources with $\rho$ > 4.0\,mas and $X_{\rho}$ > 4. Among them, 53 sources have CARMS > 0.3 and 41 sources have CARMS > 0.4. Out of the 22 sources with CARMS values $\leq$ 0.3, 20 sources have their $z$ available, and 15 sources have $z$ < 0.7. The median $z$ of these 20 sources is 0.25, which is only one fifth of the median $z$ of the 2198 sources with known $z$. A small fraction
of these sources seem to be weak but nearby optical objects.
It is important to investigate this further.

\subsection{Magnitudes of the position differences}

With an improvement in $Gaia$ position estimates in the near future, 
the number of the sources with $X_{\rho}$ > 4 
may continue to increase. However, there should be no significant increase in the number of the sources with extremely large differences, for instance $\rho$ > 4.0\,mas; currently, there are 79 sources, less than 4 percent. 

As shown in Tables \ref{magnitude_1960} and \ref{magnitude_389}, there are 615 sources with $G$ < 18 mag, and
the mean semi-major axis of the error ellipses of the $Gaia$ positions for these 615 sources is already smaller than 0.1\,mas. 
In this sample of 615 sources, 181 sources have $X_{\rho}$ > 4, 2/5. 
About 74 percent of these 181 sources have $\rho$ < 1.5\,mas; the median $\rho$ is $\sim$0.8\,mas. Therefore, the magnitude of $\rho$ for the majority of the sources with $X_{\rho}$ > 4 is expected to be at the same level as source structure effects and core shift. For the 434 sources with $X_{\rho}$ $\leq$ 4, the median $\rho$ is $\sim$0.24\,mas, which is at the same level as their uncertainties dominated by VLBI. This may provide insights on the final agreement of source positions between $Gaia$ and VLBI for the whole ensemble of common sources. 

If we assume that the median uncertainty of the $Gaia$ source positions 
at higher optical magnitudes is $\sim$0.1\,mas, which is 
better than the predicted end-of-mission accuracies but still possible \citep{2001A&A...369..339P,2014EAS....67...23D}, 
the $Gaia$ and VLBI positions will agree
with each other within their uncertainties for the 3/5 sources, 
and the median $\rho$ for these sources will be at the level of 0.24\,mas. 
There will be 2/5 sources having statistically significant position differences with a median $\rho$ of 0.8\,mas. 

Based on about 2000 evenly distributed sources over the sky with position differences of $\sim$0.24\,mas, the orientation stability of the $Gaia$ frame with respect to the ICRF3 may be achieved at the level of ten microarcseconds ($\mu$as); it is sufficient enough to detect systematic position differences between $Gaia$ and VLBI at the level of hundreds of $\mu$as. Several hundreds of sources with well-detected position differences at such levels will provide invaluable information to investigate the physical properties of radio sources.

\subsection{Directions of the position differences}
\label{directions}
Source structure and core shift are expected to cause the derived source positions from VLBI to shift  
towards the jets. If the VLBI-to-$Gaia$ position vectors
are opposite to the directions of the radio jets, as shown for the source 1928$+$738 in the 
bottom-right panel of Fig. \ref{mojave}, the position differences can be 
explained by source structure effects or core shift. However, it seems to be difficult to explain these position vectors
along the jets, as shown for the source 1803$+$784 in the bottom-left panel, by the effects of radio 
source structure and core shift.
The recent studies have demonstrated that 
the VLBI-to-$Gaia$ position vectors favor the directions both along and opposite to the 
jets \citep{2017A&A...598L...1K,2019MNRAS.482.3023P}, and more sources have these position
vectors along the jets than opposite to the jets. The presence of parsec-scale optical jet 
structure in the directions of radio jets is proposed to explain the phenomenon in these studies.

\begin{figure*}
   \centering
        \includegraphics[width=0.69\textwidth]{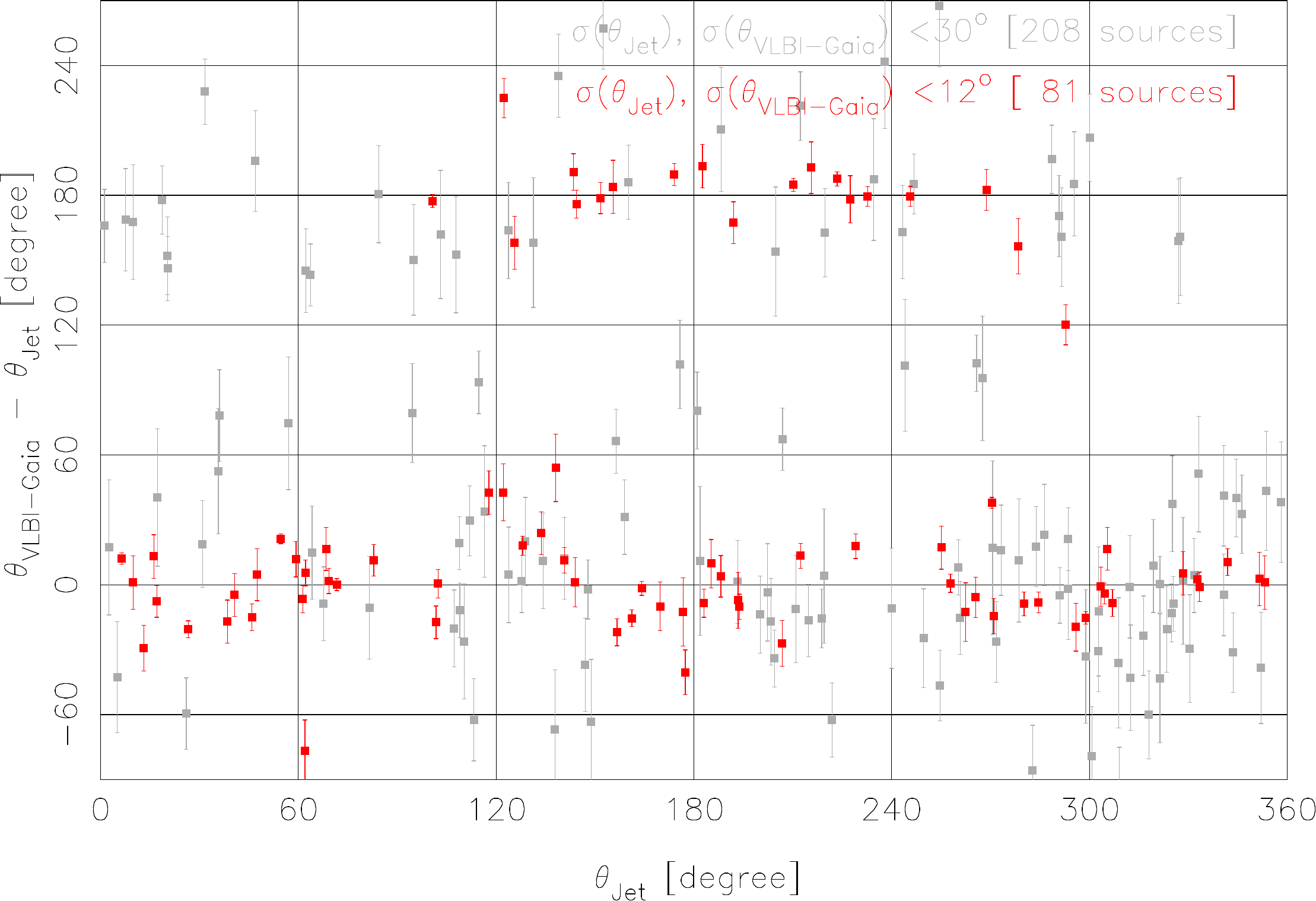}
   \caption{Angles of the VLBI-to-$Gaia$ position vectors with respect to the jet directions as a function of the jet position angles based on the MOJAVE data. The error bars shown are the combined uncertainties from the formal errors of the two directions. There are 327 sources with robust multi-epoch and multi-jet position angles, cross-matched from the 3142 sources. Out of them, 208  sources have both the uncertainties of the VLBI-to-$Gaia$ position directions and the median jet directions smaller than 30 degrees and are shown as gray dots. There are 81 sources with those uncertainties smaller than 12 degrees, shown as red dots. For these 81 sources, the median $\rho$ is 0.93\,mas, and the $X_{\rho}$ values are larger than 3.3. Among them, 54 sources have the directions of the position differences along the jet directions within 25 degrees and their $\rho$ are in the range 0.2--28.0\,mas; 17 sources have the directions of the position differences opposite to the jet directions and their $\rho$ are in the range 0.2--39.1\,mas.}
              \label{mojave_jet_angles}%
    \end{figure*}

We compared the directions of the VLBI-to-$Gaia$ position vectors and of 
the radio jets based on the MOJAVE data. The jet directions were calculated as the median values of the jet position angles for 
the multiple jets of each individual source
in the MOJAVE project. These jet position angles were robustly determined from multiple-epoch 
measurements by MOJAVE \citep{2018ApJS..234...12L}. Figure \ref{mojave_jet_angles} shows the 208 sources with the uncertainties of both the jet position angles and the VLBI-to-$Gaia$ position directions smaller than 30 degrees in gray dots and
the 81 sources with those uncertainties smaller than 12 degrees in red dots.
About 88 percent of these 81 sources have the VLBI-to-$Gaia$ position vectors parallel to
the jet directions within 25 degrees and 96 percent within 45 degrees. 
It enhances the already known results from \citet{2017A&A...598L...1K} and \citet{2019MNRAS.482.3023P}
with stronger evidence. The majority of the sources have the directions of the position
vectors along the jets and a significant fraction of sources have those vectors opposite to the jets, 
also confirmed by this small sample of well-determined jet position angles. 

\begin{figure*}
   \centering
        \includegraphics[width=0.49\textwidth]{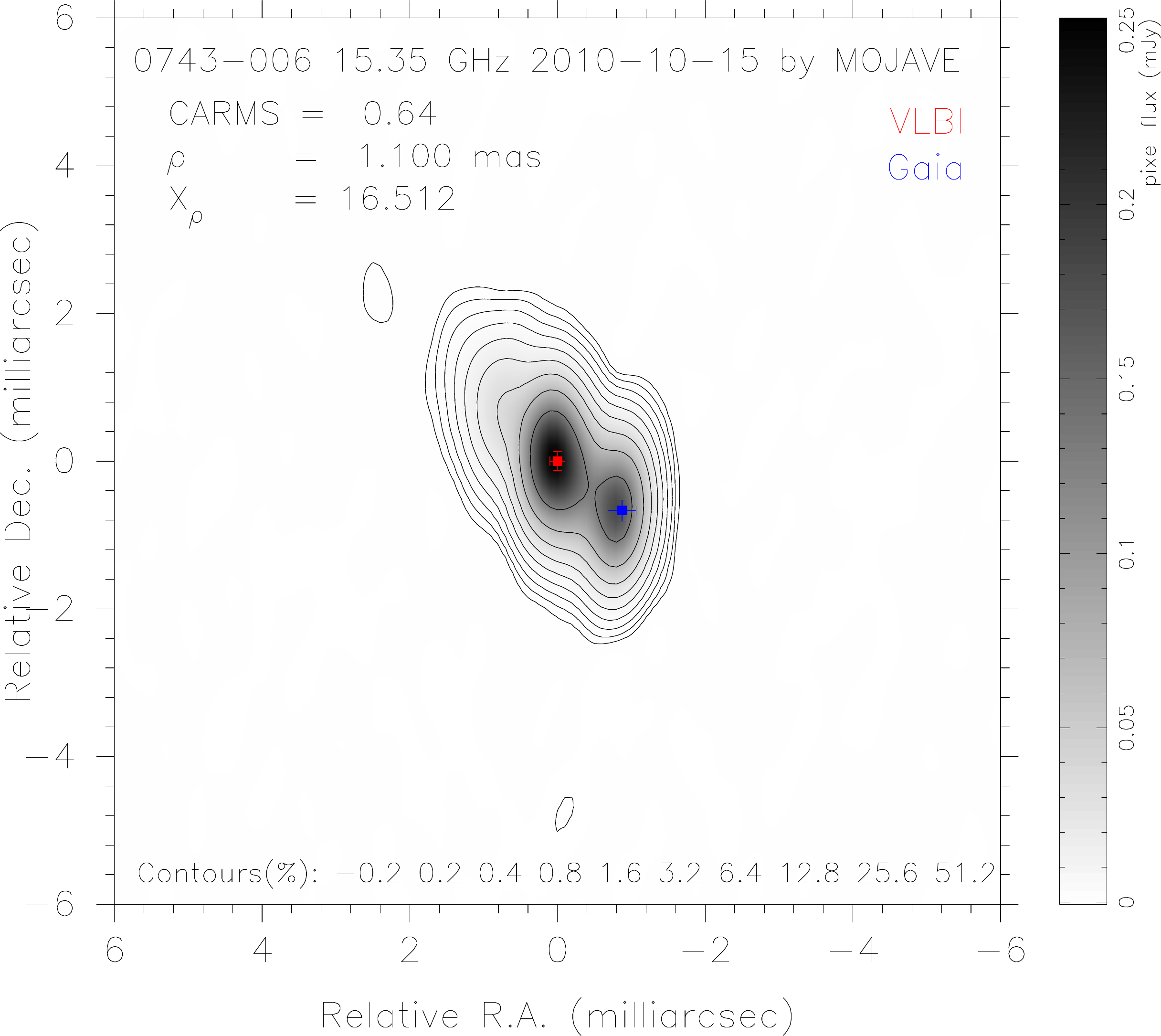}
                \includegraphics[width=0.49\textwidth]{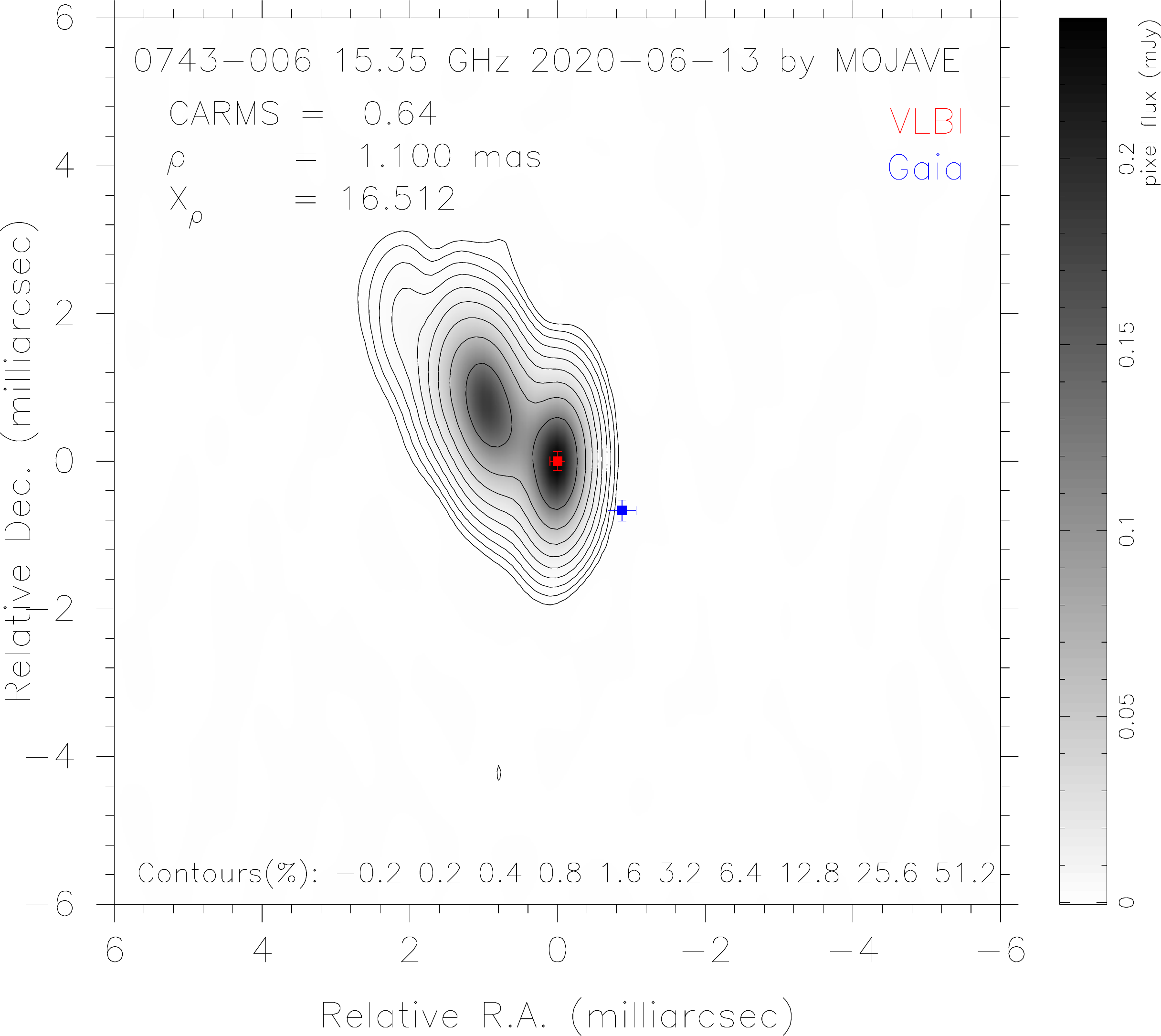}
   \caption{MOJAVE images of source 0743$-$006 (CARMS=0.64) 
at 15 Oct. 2010 (left) and at 13 Jun. 2020 (right). See the caption of Fig. \ref{mojave} for the plot design. 
The peak of flux changed between the two components from 2010 to 2020, as indicated by the red dots. Based on its 
jet motions from model fitting, it was suggested in the MOJAVE project that the source has two-sided jets and the core is located close to the component marked as the red dot in the left plot.
It seems to be possible that the south-west component is the core, meaning that the source has a one-sided jet.}
              \label{mojave_0743}%
    \end{figure*}

However, we address several cases where the jet position angles
can be determined in the opposite direction. Figure 
\ref{mojave_0743} shows the images of source 0743$-$006, one of the ICRF3 defining sources but with
the CARMS value of 0.64, at two different epochs. It has two compact components separated by 
$\sim$1\,mas and a fuzzy emission region extending to the north-east direction. The peak of flux changed between the two components from 2010 to 2020. According to its jet motions from model fitting, which are relatively small and weak for this particular source, the core was suggested by MOJAVE to be the peak of flux in the image from 2010, 
and consequently it has two-sided jets. 
It seems that the south-west component can be the core, which means that 
the source actually has a one-sided jet. In this case, the jet position angle
can be determined with an offset of 180 degrees. The source has $\rho$=1.1\,mas and $X_{\rho}$=16.5. 
If the south-west component is the core, the difference between its $Gaia$ and VLBI positions can be 
explained by its radio source structure. As we can see, in the
right-hand plot, if we move the VLBI position to the next component to
the upper-left, then the $Gaia$ position fits very well the core. 

\begin{figure*}
   \centering
        \includegraphics[width=0.49\textwidth]{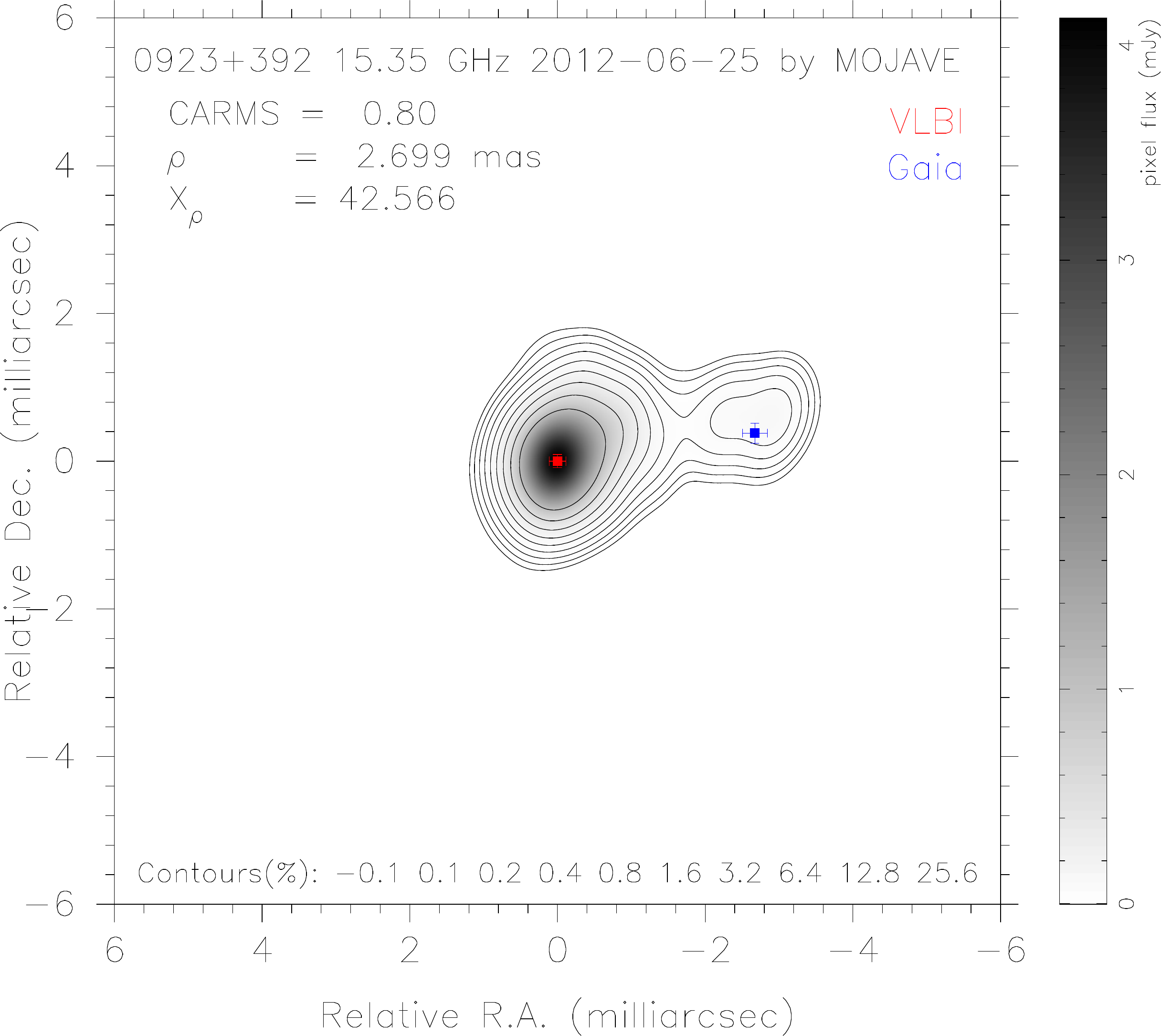}
                \includegraphics[width=0.49\textwidth]{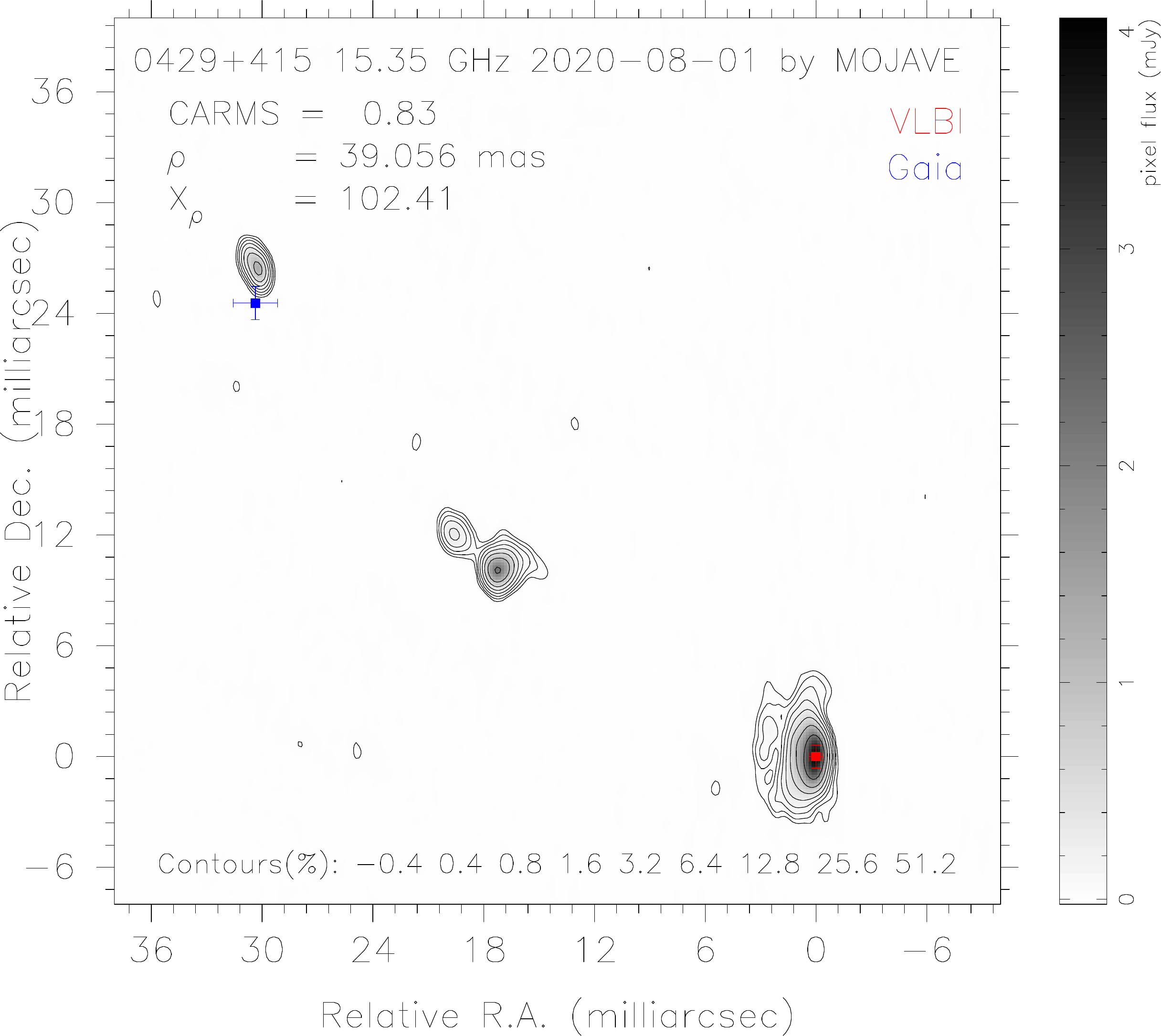}
   \caption{Explanation of the large $Gaia$-VLBI position differences for two sources, 0923+392 with $\rho$=2.7\,mas (4C39.25, CARMS=0.80, left) and 0429+415 with $\rho$=39.1\,mas (CARMS=0.83, right), based on the MOJAVE images. See the caption of Fig. \ref{mojave} for the plot design. According to the spectral index images from the MOJAVE project, the cores are not the brightest components in the images. The core of the source 0923+392 is the western, weak component, and the core of the source 0429+415 is the north-east component. Their $Gaia$ positions are located close to the cores, given that the VLBI positions are located at the peaks of flux. These two sources strongly demonstrate the effects of source structure on the position differences between VLBI and $Gaia$ --- the source positions from geodetic VLBI are dominated by the positions of the peak fluxes, whereas the optical positions from $Gaia$ are located close to the cores.}
              \label{mojave_0923+392}%
    \end{figure*}

We further discuss two cases, sources 0923$+$392 and 0429+415, of extremely large and statistically significant position differences between VLBI and $Gaia$, which can be explained by their radio structure. Their MOJAVE images are shown in Fig. \ref{mojave_0923+392} with their relative positions between VLBI and $Gaia$ illustrated. The figure demonstrates that the source positions from geodetic VLBI are dominated by the positions of the peak fluxes, whereas the optical positions from $Gaia$ are located close to the cores. The separations between the cores and the jets for the CRF sources are typically at the mas level as demonstrated in Figs. \ref{mojave} and \ref{mojave_0743} and up to tens of mas as shown in Fig. \ref{mojave_0923+392}. We should emphasize that for a significant number of sources the VLBI position seems to be that of a jet component rather than the core.
Without absolute position information in the MOJAVE images, however, 
we have no knowledge of where the VLBI position really is. Since the VLBI position 
to the core in the MOJAVE images is so large if it locates at different jet 
components for the cases like these two sources, 
phase referencing observations can determine the positions 
of the jet components with sufficient accuracy, which will allow us to 
locate the VLBI position within the image. This will eventually help to understand where the $Gaia$ position locates.
One also should notice from Fig. \ref{mojave_0923+392} that since the cores of these two sources are not the brightest components, without spectral index images it will be difficult to identify them from radio images, which can lead to a shift of 180 degrees in determining jet position angles.

To conclude, our study suggests that radio source structure is one of the major factors causing the 
position differences and that the optical jet structure tends to be also strong for the sources with extended structure at cm-wavelengths.

\section{Conclusion}

   We made the conclusion based on the position differences between the $Gaia$ EDR3 and the ICRF3 as follows:
   \begin{enumerate}
      \item The arc lengths $\rho$ of the $Gaia$ and VLBI position differences increase with the CARMS values.
      \item The majority of the sources with statistically significant arc lengths, $X_{\rho}$ > 4, are associated with the extended sources. For instance, the median CARMS of the 432 sources with $X_{\rho}$ > 4 is 0.42, while that of the remaining 2028 sources is only 0.22. 
      \item For the sources with $\rho$ > 4.0\,mas and $X_{\rho}$ > 4, the majority, 70 percent, have extended structure. The source 0429+415 has been used as an example to demonstrate that based on the MOJVAE image shown in Fig. \ref{mojave_0923+392}.
      \item Distinct relations between the optical magnitudes and the redshifts are found for the sources with and without statistically significant position differences. The sources with $X_{\rho}$ > 4 have substantially smaller redshift values, $\sim$0.3. Our study suggests that a small fraction of these sources may be associated with the weak but 
      nearby (small redshifts) optical objects.
\item We argue that core shift can contribute to the position differences if the source has extended structure.
\item The $Gaia$ and VLBI position differences can be well explained through the radio images for several sources as examples. The vectors of the $Gaia$ and VLBI position differences are parallel to the radio-jet directions, which is confirmed with stronger evidence. 
   \end{enumerate}

\begin{acknowledgements}
We would like to thank the reviewer Fran\c{c}ois Mignard for his helpful comments. This research has made use of data from the MOJAVE database that is maintained by the MOJAVE team \citep{2018ApJS..234...12L}.   
All components of the International VLBI Service for Geodesy and Astrometry are deeply appreciated for providing the VLBI observations. This research was supported by the Academy of Finland project No. 315721 and the
National Natural Science Foundation of China No. 11973023. SL is supported by the DFG grant No. HE5937$\/$2-2.
\end{acknowledgements}

%
%
%
%
%
%
%
%
%
%
%
%
\bibliographystyle{aa} 
\bibliography{gaia_crf} 

\end{document}